\documentclass[11pt,a4paper]{article}
\pdfoutput=1
\usepackage{setspace}
\usepackage{jheppub}
\usepackage{rotating}
\usepackage{array}
\usepackage{amsmath}
\usepackage[normalem]{ulem}
\usepackage{slashed}
\usepackage{booktabs}
\usepackage[pdftex,table,dvipsnames]{xcolor}
\usepackage{units}
\usepackage{multirow}
\usepackage{comment}
\usepackage{url}
\usepackage{xspace}
\usepackage{color}
\usepackage{float}
\usepackage{lipsum}
\usepackage[utf8]{inputenc}
\usepackage{bbold}
\DeclareUnicodeCharacter{2032}{\ensuremath{^{\prime}}}
\preprint{TTK-20-40, P3H-20-068}

\title {On the challenges of searching for GeV-scale long-lived particles at the LHC}

\author{Elias~Bernreuther,}
\author{Juliana~Carrasco~Mejia,}
\author{Felix~Kahlhoefer,}
\author{Michael~Kr\"{a}mer}
\author{and Patrick~Tunney}

\affiliation{Institute for Theoretical Particle Physics and Cosmology (TTK), RWTH Aachen University, \\ D-52056 Aachen, Germany}

\emailAdd{ebernreuther@physik.rwth-aachen.de}
\emailAdd{kahlhoefer@physik.rwth-aachen.de}
\emailAdd{mkraemer@physik.rwth-aachen.de}

\abstract{
Many models of dark matter predict long-lived particles (LLPs) that can give rise to striking signatures at the LHC. Existing searches for displaced vertices are however tailored towards heavy LLPs. In this work we show that this bias severely affects their sensitivity to LLPs with masses at the GeV scale. To illustrate this point we consider two dark sector models with light LLPs that decay hadronically: a strongly-interacting dark sector with long-lived exotic mesons, and a Higgsed dark sector with a long-lived dark Higgs boson. We study the sensitivity of an existing ATLAS search for displaced vertices and missing energy in these two models and find that current track and vertex cuts result in very low efficiency for light LLPs. To close this gap in the current search programme we suggest two possible modifications of the vertex reconstruction and the analysis cuts. We calculate projected exclusion limits for these modifications and show that they greatly enhance the sensitivity to LLPs with low mass or short decay lengths.}


\begin{document}

\notoc

\maketitle

\flushbottom

\section{Introduction}

While modern particle physics has established the Standard Model (SM) as an accurate description of all known particles and their interactions, modern cosmology tells us that only a small fraction of the energy density of the universe can be explained in this way. Given the complexity of visible matter, we should expect that there exist other \emph{dark} sectors containing a rich internal structure with stable and unstable states and new interactions. Interest in such dark sectors has risen sharply in recent years as a result of the non-observation of either new physics at the LHC or evidence for simpler dark matter (DM) models in laboratory experiments.

The most exciting feature of dark sectors is that, in addition to accounting for some or all of the DM density in the universe, they predict novel signatures that can be searched for at existing or near-future experiments (see e.g.\ ref.~\cite{Beacham:2019nyx}). Of particular interest is the generic existence of long-lived particles (LLPs) in many dark sectors, which can give rise to displaced decays at accelerator experiments~\cite{Co:2015pka,Izaguirre:2015zva,Davoli:2017swj,Davoli:2018mau,Berlin:2018jbm,Calibbi:2018fqf,Belanger:2018sti,No:2019gvl,Duerr:2019dmv}. Indeed, the search for LLPs at the LHC is now rapidly gathering pace and a broad and comprehensive search programme is being defined~\cite{Lee:2018pag,Alimena:2019zri}.

The purpose of the present work is to point out a gap in the existing set of searches for LLPs and suggest strategies for closing it. For this purpose we consider two examples for dark sectors containing GeV-scale LLPs, which are largely unconstrained by existing analyses. The first example is a strongly interacting dark sector that confines at low energies~\cite{Strassler:2006im,Hochberg:2014kqa,Kribs:2016cew,Bernreuther:2019pfb}. At the LHC such a dark sector would reveal itself in the form of dark showers containing both stable (i.e.\ invisible) and long-lived dark mesons~\cite{Cohen:2017pzm,Pierce:2017taw,Beauchesne:2017yhh,Renner:2018fhh}. The second example is a Higgsed dark sector, in which the masses of the various particles are generated via spontaneous symmetry breaking~\cite{Duerr:2016tmh,Bell:2016uhg}. If the dark Higgs boson is the lightest particle in the dark sector, it can be readily produced in association with other dark sector states~\cite{Duerr:2017uap,Aad:2020sef} and travel macroscopic distances before decaying into SM particles~\cite{Darme:2017glc}.

Although both of these dark sectors can in principle be realised across a wide range of mass scales, we will be interested in dark sectors that also contain a viable DM candidate. This requirement motivates us to focus on LLPs with masses in the range 10--100 GeV.
Since these particles are too heavy to be produced at $B$ factories or in fixed-target experiments, the LHC offers a unique opportunity for exploring these models. Existing searches for LLPs are however often influenced by the expectation of new physics at the TeV scale and may not be ideally suited to searching for lighter LLPs.

As a specific example we consider searches for displaced vertices (DVs) in association with missing energy (MET), which exploit the fact that in dark sector models LLPs are typically produced in association with DM candidates or other invisible particles. A first analysis of this signature has been performed by the ATLAS collaboration in ref.~\cite{Aaboud:2017iio}.\footnote{Similar searches have been performed by CMS in refs.~\cite{Sirunyan:2018njd,Sirunyan:2019nfw}. However, these searches target TeV-scale gluinos as LLPs and the available information on the vertex reconstruction efficiencies is insufficient for a reinterpretation in our context.} In this analysis an accurate reconstruction of the DV mass is essential for background suppression. Unfortunately, such a reconstruction is challenging if the decaying particle is highly boosted, such that most of the decay products have small impact parameter (i.e.\ their tracks point back towards the interaction point). We show that removing charged tracks with small impact parameter introduces a substantial bias in the reconstructed vertex mass, which severely affects the sensitivity of these searches to GeV-scale LLPs.

In this work we propose various modifications of the existing analysis strategy to enhance the sensitivity of searches for DVs and MET to low-mass dark sectors. We show that relaxed analysis cuts as well as an improved reconstruction of the DV mass can substantially enhance the sensitivity of these searches. A related study has been done in ref.~\cite{Cottin:2018kmq} in the context of searches for heavy neutral leptons in the DV+lepton channel. The present work goes beyond this analysis in that we consider not only modifications to the analysis cuts but also to the vertex reconstruction, as well as more realistic detector efficiencies. Including these modifications in future analyses will make it possible to probe parameter regions of the dark sector models that contain viable DM candidates but are challenging to constrain with any other search strategy, opening up a new direction for DM searches at the LHC.

The remainder of this paper is structured as follows. In section~\ref{sec:dark_sectors} we introduce the two dark sectors that we investigate, pointing out their common features and differences and identifying the relevant LLPs. We then discuss the existing search for such dark sectors in section~\ref{sec:analysis}, identifying the shortcomings of the current approach and possible ways to address them. We present our analysis of existing constraints and projected sensitivities in section~\ref{sec:results} before concluding in section~\ref{sec:conclusions}.

\section{Model Details}
\label{sec:dark_sectors}

In this section we introduce two different models of dark sectors containing stable DM candidates as well as unstable states at the GeV scale: a strongly interacting dark sector and a Higgsed dark sector. Both of these models predict LHC events with a combination of invisible and long-lived particles in the final state, leading to the characteristic signature of missing energy together with a displaced vertex, which will be the focus of our analysis below.

For both cases we assume that the interactions between the dark sector and SM particles are mediated by a $Z'$ vector boson arising from a new $U(1)'$ gauge group under which both dark sector and SM particles are charged. Although couplings to other SM particles may in general be present (and are typically required by anomaly cancellation~\cite{Ellis:2017tkh,Caron:2018yzp,ElHedri:2018cdm}), we focus on the coupling of the $Z'$ to quarks, which allows for dark sector particles to be produced at hadron colliders. In other words, we extend the SM Lagrangian by
\begin{equation}
 \mathcal{L}_{Z'} = \frac{1}{2} m_{Z'}^2 Z'^\mu Z'_\mu + g_q \sum_q Z'^\mu \bar{q} \gamma_\mu q \; ,
\end{equation}
where we have assumed that the $Z'$ couples with equal strength to all quarks. More general coupling structures are not expected to qualitatively change the phenomenology discussed below. Likewise, the precise value of the $Z'$ mass is irrelevant for the subsequent discussion and will therefore be fixed to $m_{Z'} = 1 \, \mathrm{TeV}$ in what follows. The quark coupling $g_q$, on the other hand, determines not only the production cross section for dark sector states, but also their decay width back into SM particles, which is decisive for the collider phenomenology. Increasing $g_q$ enhances all cross sections but reduces the decay length of dark sector states. Searches for displaced decays are therefore typically sensitive to a finite range of couplings, typically around $g_q \sim \mathcal{O}(0.01)$.

The two dark sectors that we consider furthermore have in common that the DM relic density is set by interactions internal to the dark sector, which are independent of the value of $g_q$. In other words, for appropriate choices of dark sector parameters, the observed DM relic density can be reproduced for any value of $g_q$. The only requirement is that the interactions between the dark sector and the SM are strong enough to establish thermal equilibrium between the two sectors, which is always satisfied for couplings that can be probed at the LHC~\cite{Kahlhoefer:2018xxo}. We will therefore treat $g_q$ as a free parameter in the following.

\subsection{Strongly-interacting dark sector}

Following ref.~\cite{Bernreuther:2019pfb}, we consider a dark sector consisting of two flavours of dark quarks $q_\mathrm{d}$ with equal mass $m_q$ that transform in the fundamental representation of a new $SU(3)_\mathrm{d}$ gauge group and also carry opposite charge $\pm e_\mathrm{d}$ under the $U(1)'$:
\begin{align}
\label{eq:lagrangian_massless}
\mathcal{L}_1 = -\frac{1}{4}F_{\mu\nu}^a F^{\mu\nu a} + \sum_{i=1}^2 \overline{q}_{\mathrm{d},i} (i \slashed{D} - m_q) q_{\mathrm{d},i} - e_\mathrm{d} Z^\prime_\mu \left(\overline{q}_{\mathrm{d},1}\gamma^\mu q_{\mathrm{d},1}- \overline{q}_{\mathrm{d},2}\gamma^\mu q_{\mathrm{d},2}\right) \; .
\end{align}
In analogy to QCD the dark sector confines below some scale $\Lambda_\mathrm{d}$ giving rise to three dark pions with $U(1)'$ charge $2 e_\mathrm{d}$, 0 and $-2 e_\mathrm{d}$. All three dark pions can be stabilised by a suitable symmetry and therefore constitute the DM candidates in this set-up~\cite{Bernreuther:2019pfb}. Specifically, for the dark pions to be stable it is required that the square of the dark quark charge matrix $Q^2 \propto \mathbb{1}$ to avoid the decay $\pi_\mathrm{d} \to Z^{\prime \ast}Z^{\prime \ast}$ through the anomaly triangle diagram (analogous to the decay of SM pions into photons). Moreover, imposing a dark $G$-parity~\cite{Bai:2010qg, Berlin:2018tvf} that forbids all dark pion decays to SM particles additionally requires an even number of dark quark flavours, e.g. $N_f=2$ in the scenario considered here.

Heavier mesons, in particular the vector mesons analogous to the SM $\rho$ meson, are in general unstable against decays~\cite{Berlin:2018tvf}. Of particular interest is the vector meson neutral under the $U(1)'$ called $\rho_\mathrm{d}^0$, which mixes with the $Z'$ meson and thereby obtains couplings to SM quarks. For $m_{\rho_\mathrm{d}} < 2 m_{\pi_\mathrm{d}}$ the $\rho_\mathrm{d}^0$ then exclusively decays into visible final states with a partial decay width given by 
\begin{equation}
\Gamma\left(\rho^0\to q \overline{q}\right) =
 \frac{1}{\pi}\frac{g_q^2e_\mathrm{d}^2}{g^2}\,m_{\rho_\mathrm{d}}\left(\frac{m_{\rho_\mathrm{d}}}{m_{Z'}}\right)^4\left(1-4\frac{m_{q}^2}{m_{\rho_\mathrm{d}}^2}\right)^{1/2} \left(1+2\frac{m_{q}^2}{m_{\rho_\mathrm{d}}^2}\right) \; ,
\end{equation}
where $g$ denotes the $\pi_\mathrm{d}$--$\rho_\mathrm{d}$ coupling. For $m_{Z^\prime} \gg m_{\rho_\mathrm{d}}$ the mixing between the two vector bosons is suppressed and hence the $\rho_\mathrm{d}^0$ lifetime can become very long even for relatively large couplings $g_q$ and $e_\mathrm{d}$.
For example, for $m_{Z'} = 1\,\mathrm{TeV}$, $g = 3$ and $e_\mathrm{d} = 2 g_q$ we obtain
\begin{equation}
c \tau_{\rho_\mathrm{d}} \approx 4.4 \, \mathrm{mm} \times \left(\frac{m_{\rho_\mathrm{d}}}{40\,\mathrm{GeV}}\right)^{-5} \left(\frac{g_q}{0.005}\right)^{-4} \; .\end{equation}

\begin{figure}[t]
\centering
\includegraphics[width = 0.6 \textwidth]{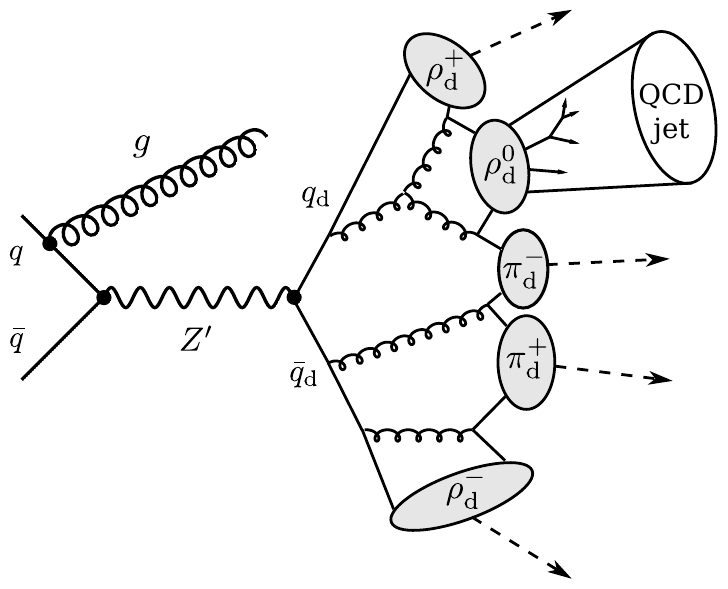}
\caption{Diagram of typical LHC signal event containing a dark shower. The $\rho_\mathrm{d}^0$ decays into SM quarks, while the other mesons escape from the detector as missing energy. Figure taken from ref.~\cite{Bernreuther:2019pfb}.} \label{fig:DarkShower}
\end{figure}

We note in passing that in the set-up introduced above the relic density of dark pions is determined by the conversion processes $\pi_\mathrm{d} \pi_\mathrm{d} \to \rho_\mathrm{d} \rho_\mathrm{d}$, which depend on the mass difference between the two types of mesons and their interaction strength $g$. For a given $\rho_\mathrm{d}^0$ mass it is therefore always possible to find values of $g$ and $m_{\pi_\mathrm{d}}$ that reproduce the observed DM relic abundance (see ref.~\cite{Bernreuther:2019pfb} for details). In the following we fix $g = 3$ and set $m_{\pi_\mathrm{d}}$ accordingly.

\begin{figure}[t]
\centering
\includegraphics[width = 0.495 \textwidth]{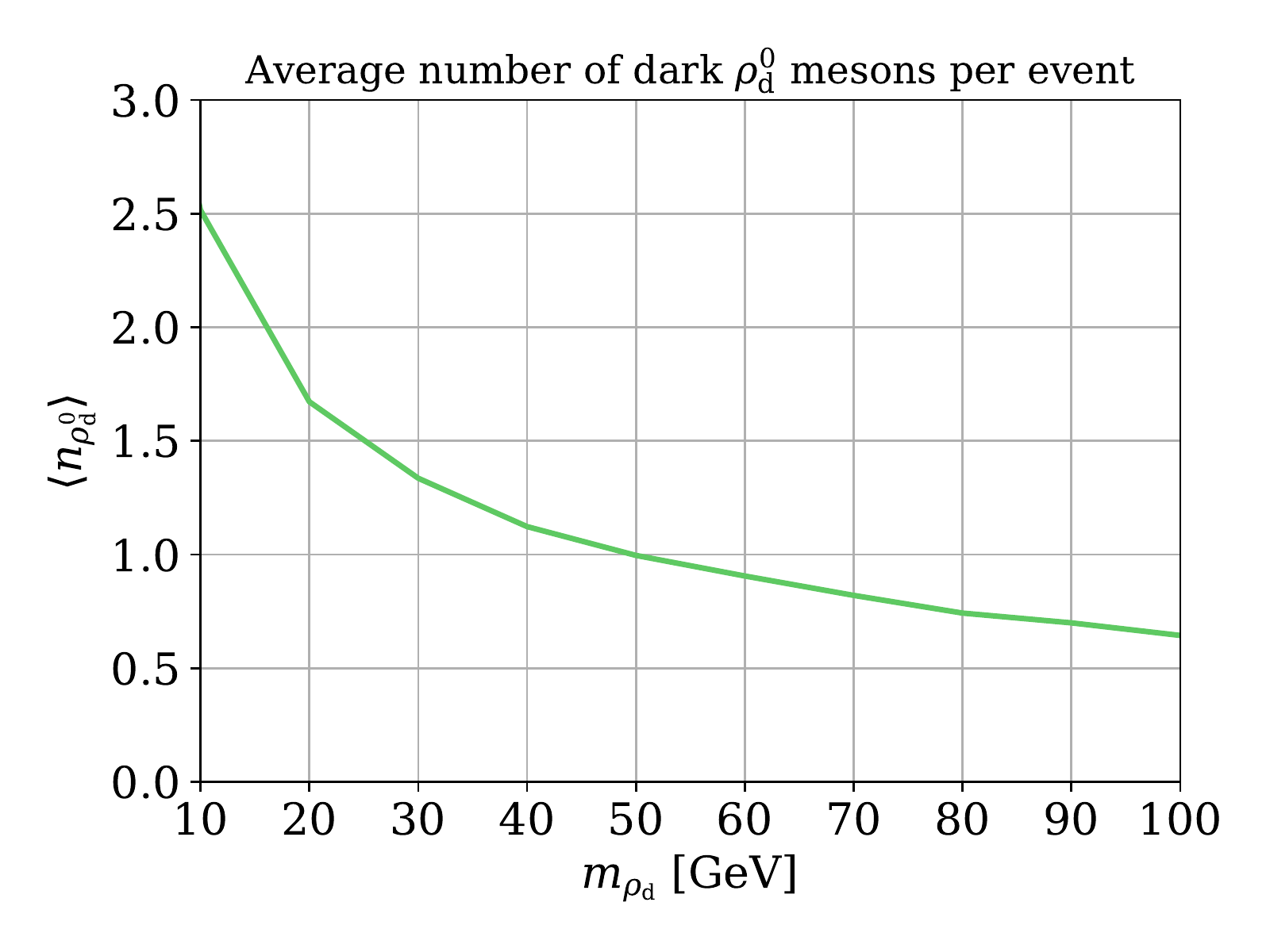}
\includegraphics[width = 0.495 \textwidth]{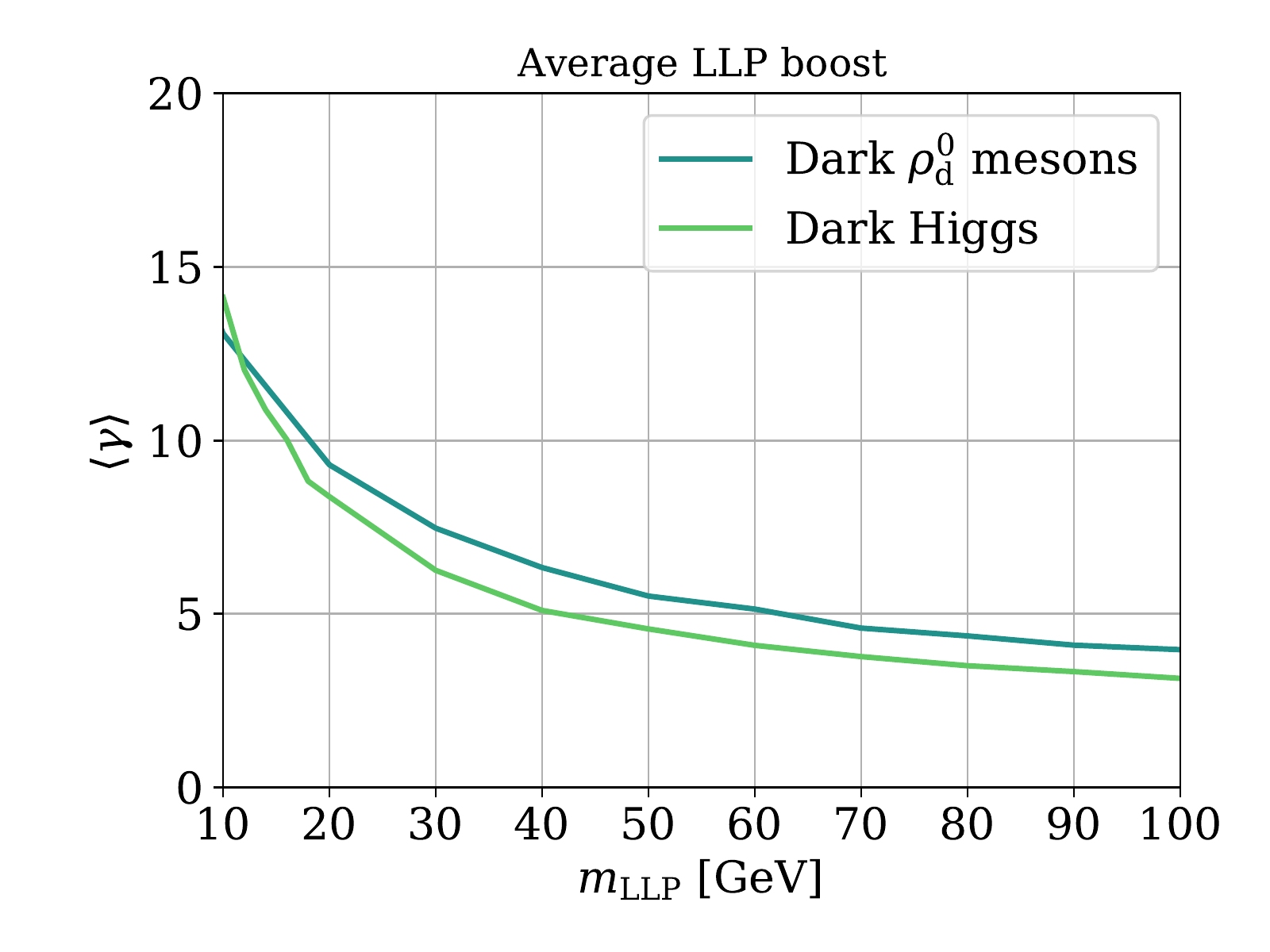}
\caption{Left: Average multiplicity of dark $\rho^0_\mathrm{d}$ mesons in a dark shower as a function of the dark rho mass. Right: Average boosts of $\rho^0_\mathrm{d}$ mesons and dark Higgs bosons as a function of their mass. The $Z'$ mass is set to 1~TeV.} \label{fig:Multiplicity} 
\end{figure}

If dark quarks are produced at the LHC through the $Z'$ mediator, they will hadronise into the various dark mesons, leading to a dark shower (see figure~\ref{fig:DarkShower}). Most dark mesons are either completely stable or extremely long-lived (leading to missing energy), but any $\rho_\mathrm{d}^0$ produced will decay into SM quarks, which immediately hadronise. Depending on the $\rho_\mathrm{d}^0$ lifetime, the resulting jet will either be prompt (such that the dark shower will appear as a semi-visible jet~\cite{Cohen:2015toa}, see also refs.~\cite{Bernreuther:2020vhm,Kar:2020bws} for recent analyses) or originate from a displaced vertex (a so-called emerging jet~\cite{Schwaller:2015gea,Sirunyan:2018njd}). For our benchmark choice $m_{Z'} = 1\,\mathrm{TeV}$ and a $\rho_\mathrm{d}^0$ mass of 40~GeV, a typical event contains either one or two $\rho_\mathrm{d}^0$ meson with an average boost of around $\gamma \approx 6$. Both the $\rho_\mathrm{d}^0$ multiplicity and the average boost decrease with increasing $\rho_\mathrm{d}^0$ mass (see figure~\ref{fig:Multiplicity}).

\subsection{Higgsed dark sector}
\label{sec:darkhiggs}

Following ref.~\cite{Duerr:2016tmh}, we consider a Majorana DM particle $\chi$ coupled to a complex scalar field $S$ with Yukawa coupling strength $y_\chi$. The scalar field obtains a vacuum expectation value $w$, which generates the DM mass $m_\chi = y_\chi w / \sqrt{2}$. We furthermore assume that the scalar and the DM particle are charged under the $U(1)'$, such that the vacuum expectation value of $S$ spontaneously breaks the gauge symmetry  and generates the $Z'$ mass $m_{Z'}$ via the Higgs mechanism.\footnote{As discussed in Ref.~\cite{Duerr:2016tmh}, this set-up is a simplified version of models for DM from gauged baryon number~\cite{Duerr:2013dza,Duerr:2013lka,Duerr:2014wra} in the sense that we do not consider the additional fermions which are necessary for anomaly cancellation, but which can be sufficiently heavy to be irrelevant for the phenomenology discussed here.} Writing $S = (s + w)/\sqrt{2}$, where $s$ denotes the dark Higgs boson, the interaction Lagrangian is given by
\begin{equation}
 \mathcal{L}_2 = - \frac{1}{2} g_\chi Z^{\prime \mu} \bar{\chi} \gamma^5 \gamma_\mu \chi - \frac{y_\chi}{2 \sqrt{2}} s \bar{\chi} \chi + 2 \, g_\chi \, Z^{\prime \mu} Z^\prime_\mu \left( g_\chi \, s^2 + m_{Z'} s \right) \; ,
\label{eq:Ldark}
\end{equation} 
where $g_\chi = y_\chi m_Z' / (2 \sqrt{2} m_\chi)$. The dark sector is hence fully characterised by $m_\chi$, $m_s$ and $y_\chi$ (or equivalently $g_\chi$), as well as the $Z'$ mass $m_{Z'}$.

In principle the dark Higgs boson can mix with the SM Higgs boson and thereby obtain couplings to SM fermions. Here we instead assume that this mixing is negligible, and decays into SM fermions proceed dominantly via a loop involving two $Z'$ bosons~\cite{Darme:2017glc}. The resulting decay width of the dark Higgs is then given by~\cite{Batell:2009yf}
\begin{equation}
 \Gamma\left(s\to q \overline{q}\right) = \frac{3 \, g_q^4 \, g_\chi^2 \, m_s}{32\pi^5} \frac{m_q^2}{m_{Z'}^2} \left(1-\frac{4 m_q^2}{m_{s}^2}\right)^{3/2} \left| I\left(\frac{m_s^2}{m_{Z'}^2},\frac{m_q}{m_{Z'}^2}\right) \right|^2 \; ,
\end{equation}
where
\begin{equation}
 I(x_s, x_q) \equiv \int_0^1 \mathrm{d}y \int_0^{1-y} \mathrm{d}z \frac{2-(y+z)}{(y+z)+(1-y-z)^2 x_q - y z x_s} \approx \frac{3}{2}
\end{equation}
for $m_{Z'} \gg m_s, m_q$. Note that even though the $Z'$ is assumed to couple uniformly to all quarks, the loop-induced decay width is helicity suppressed and hence proportional to $m_q^2$. Because of the loop suppression and the strong dependence on $g_q$, the decay width of the dark Higgs boson can be very small, corresponding to macroscopic decay lengths. For example, for $m_{Z'} = 1 \, \mathrm{TeV}$ and $g_\chi = 1.2$, we find
\begin{equation}
 c \tau_s \approx 2.3 \, \mathrm{cm} \times \left(\frac{m_s}{40\,\mathrm{GeV}}\right)^{-1} \left(\frac{g_q}{0.01}\right)^{-4} \; .
\end{equation}

We note that for $m_s < m_\chi$ the DM relic density is determined by the annihilation process $\chi \chi \to s s$. Hence, for a given value of $m_\chi$ it is possible to determine $g_\chi$ such that the observed DM relic abundance is reproduced. Since the precise value of the DM mass is irrelevant for our analysis, we fix $m_\chi = 200\,\mathrm{GeV}$ in the following, which then implies $g_\chi = 1.2$ (see ref.~\cite{Duerr:2016tmh} for details).

\begin{figure}[t]
\centering
\includegraphics[clip,trim=30 570 100 30,width = 0.9\textwidth]{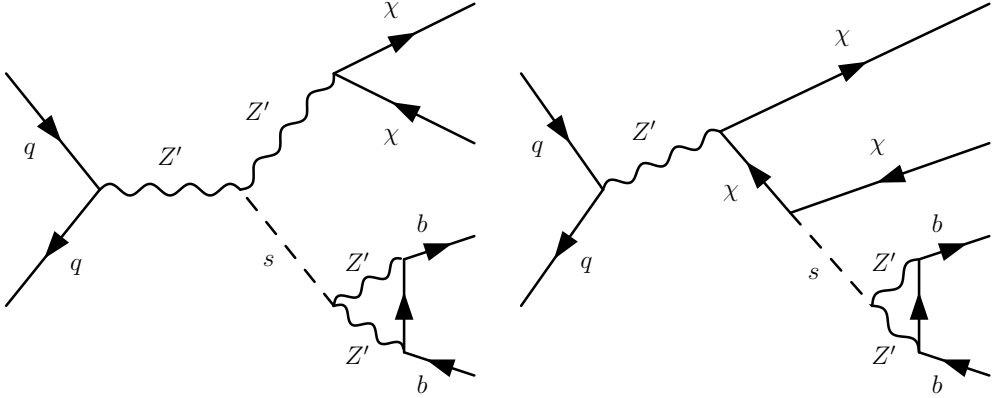}
\caption{Examples of typical processes producing a dark Higgs boson together with a pair of DM particles. The dark Higgs boson subsequently undergoes loop-induced decays into bottom quarks.} \label{fig:DarkHiggs}
\end{figure}

At the LHC dark Higgs bosons can be produced either via final state radiation from a DM particle or via Higgs-strahlung from an intermediate $Z'$ boson (see figure~\ref{fig:DarkHiggs}). As pointed out in ref.~\cite{Duerr:2017uap}, the resulting combination of visible and invisible particles in the final state automatically leads to sizeable missing energy, even if there is no hard jet from initial state radiation. For the parameters that we consider, the fraction of DM events with $\slashed{E}_T > 200 \, \mathrm{GeV}$ that contain a dark Higgs boson is approximately 37\%. Dark Higgs bosons produced in this way have a typical boost of $\gamma \approx 5$ for $m_s = 40\,\mathrm{GeV}$, which again increases rapidly for decreasing dark Higgs mass (see right panel of figure~\ref{fig:Multiplicity}).

\bigskip

Let us finally point out the differences between the two dark sector models that we consider. Most importantly, while the dark rho mesons decay with equal probability into all light quarks, the dark Higgs bosons decay dominantly into bottom quarks. The lifetimes of the LLPs furthermore depend in different ways on the portal interaction: While the dark rho meson decay width is proportional to $g_q^2 m_{\rho_\mathrm{d}}^5 / m_{Z'}^4$, the dark Higgs boson decay width is proportional to $g_q^4 m_s / m_{Z'}^2$. These different scalings reflect the different origins of the long lifetime, i.e.\ suppressed mass mixing in the former case and loop-induced decays in the latter. Finally, while in strongly-interacting dark sectors there can be several displaced vertices in the same event, in the second model the probability of producing multiple dark Higgs bosons is more strongly suppressed. In the following we will see how these difference affect the LHC sensitivity for dark sectors with light LLPs.

\section{LHC searches for displaced vertices and missing energy}
\label{sec:analysis}

Having introduced the two dark sector models that we want to study, we now turn to the reinterpretation of an existing ATLAS search for DVs and MET based on an integrated luminosity of $32.8\,\mathrm{fb^{-1}}$~\cite{Aaboud:2017iio}. This search targets DVs with a large number of tracks and is therefore particularly sensitive to LLPs decaying into SM hadrons. Moreover, the ATLAS collaboration has provided extensive additional material, such as efficiency maps, that can be used for recasting this analysis. However, as we will see below, the analysis in its current form is best suited to searching for heavy resonances and loses sensitivity for LLPs at the GeV scale.

\subsection{Existing ATLAS analysis}
\label{sec:atlas_analysis}

The ATLAS search imposes separate requirements on track properties, vertex properties and event-level information. Tracks are included if they fulfil the following requirements:
\begin{itemize}
\item The track originates from a stable and charged particle.
\item The track has transverse momentum $p_T > 1$~GeV.
\item The track has a transverse impact parameter $d_0 > 2$ mm.
\end{itemize}
Here the transverse impact parameter is defined as the shortest distance between beam axis and the trajectory of the track.

Tracks that satisfy the properties listed above can then be used to reconstruct vertices. A DV is defined by the following criteria:
\begin{itemize}
\item Its position satisfies $4\,\mathrm{mm} < R < 300\,\mathrm{mm}$ and $|z| < 300 \, \mathrm{mm}$, where $R = \sqrt{x^2 + y^2}$ is the transverse distance to the interaction point.
\item The associated number of tracks satisfies $n_{\rm tracks} \geq 5$.
\item Its mass satisfies $m_{\rm DV} \geq 10\,\mathrm{GeV}$, where the energy of each track is calculated from its three momenta, assuming its mass to be equal to that of the SM charged pion, i.e. $m_\mathrm{DV}^2 = \left(\sum\limits_\mathrm{tracks} \sqrt{\vec{p}_\mathrm{track}^2 + m_\pi^2}\right)^2 - \left(\sum\limits_\mathrm{tracks} \vec{p}_\mathrm{track}\right)^2$.
\end{itemize}
Finally, the event as a whole must have:
\begin{itemize}
\item At least one displaced vertex.
\item Missing energy $\slashed{E}_T > 200\,\mathrm{GeV}$.
\item Due to specific triggers used during parts of data-taking, 75\% of the events should have at least one jet with $p_T > 70$~GeV or at least two jets with $p_T > 25$~GeV.\footnote{The ATLAS analysis requires these to be trackless jets, i.e.\ jets where the scalar sum of the $p_T$ is less than 5~GeV for tracks with small impact parameter. The precise requirement on the impact parameter is not provided in the recasting information, but we find that it makes no difference in practice whether or not we require these jets to be trackless and therefore do not impose this requirement.} In practice we randomly decide for each generated event whether the additional trigger requirement is applied.
\end{itemize}

To determine the sensitivity of this search to our model of a strongly-interacting dark sector, we simulate parton-level $q \bar{q} \to Z^\prime \to q_\mathrm{d} \bar{q}_\mathrm{d}$ events with \textsc{MadGraph5\_aMC{@}NLO}~\cite{Madgraph}, matched with up to one extra hard jet. We then perform showering and hadronisation with the hidden valley module~\cite{Carloni:2010tw,Carloni:2011kk} of \textsc{Pythia  8}~\cite{Pythia8}, passing the $\rho_\mathrm{d}^0$ lifetime as a fixed parameter in the \textsc{Pythia} card. In the dark Higgs model we simulate the process $q \bar{q} \to s \chi \chi$ in \textsc{MadGraph5\_aMC{@}NLO}, again matched with up to one additional hard jet, and perform showering and hadronisation with \textsc{Pythia  8}. \textsc{Delphes3}~\cite{Delphes3} then performs jet clustering with \textsc{FastJet}~\cite{FastJet,FastJet_2}. Instead of scanning over the LLP lifetime, we simply rescale all position space coordinates appropriately.

Following the ATLAS reconstruction algorithm, we build vertices from tracks meeting the above requirements, merging them if they are separated by less than $1\,\mathrm{mm}$. For our dark rho mesons, which decay predominantly into light quarks, this prescription implies that tracks stemming from a single LLP decay are merged into a single vertex. However, the situation is more subtle for LLPs decaying into heavy quarks which produce long-lived SM mesons in the ensuing decay chain: The dark Higgs decays predominantly into bottom quarks, which hadronise into unstable $B$ mesons with a proper decay length of about 0.5~mm.\footnote{A small fraction of decays also produce $\Lambda_b^0$ baryons with a proper decay length of about 0.4~mm. In the following, we will refer to $B$ mesons for simplicity, but always include these additional channels in our analysis.} However, most $B$ mesons emerging from dark Higgs decays are substantially boosted and therefore travel a longer distance. For instance, we find that after the decay of a 40~GeV (100~GeV) dark Higgs more than 70~\% (80\%) of the produced $B$ mesons travel further than 1~mm before they decay. While the ATLAS collaboration does not provide publicly available information on how to reinterpret the DV+MET analysis for the case that the LLP decay produces long-lived mesons, its vertex reconstruction techniques are highly efficient at joining tracks from $B$ meson decays to the original LLP vertex~\cite{ATLAS:2019wqx}. In our reinterpretation we therefore make the optimistic assumption that all tracks that stem from the decays of $B$ mesons produced in the LLP decay can be associated with the LLP vertex. 

We note that in addition to the pair of long-lived $B$ mesons dark Higgs decays also produce a number of charged tracks originating from the LLP vertex. Hence, even if not all the tracks from the $B$ mesons are associated with the LLP vertex, there may be be enough tracks to satisfy the $n_{\rm tracks}$ requirement. However, $m_{\rm DV}$ decreases with each track not matched to the LLP vertex, making it harder for the event to pass the selection requirements. A more realistic implementation, including the efficiency for matching tracks from $B$ meson decays to the LLP vertex, is therefore expected to decrease the sensitivity of the search to dark Higgs bosons. We encourage the experimental collaborations to make this additional information available to facilitate a more accurate reinterpretation.

After applying the selection requirements discussed above, we obtain for each event the number of DVs. To each DV we then apply the vertex level efficiencies (as a function of $m_{\rm DV}$, $n_{\rm tracks}$ and $R$) given in the recasting information of~\cite{Aaboud:2017iio}, and multiply the sum with the event level efficiency (as a function of $E_T^{\rm miss}$ and $R_{\rm max}$, which denotes the largest $R$ of a truth-level vertex in an event). Multiplying the result with the signal cross section and the luminosity then gives us the expected number of vertices for a particular model parameter point. ATLAS observes 0 DVs (in agreement with the background expectation of 0.02), which means we can exclude any model parameter point that predicts more than 3 displaced vertices at 95\% C.L.

\subsection{Distribution of displaced vertex mass}

At first sight, the search strategy described above should be sensitive to LLPs with a mass as small as $10\,\mathrm{GeV}$. In practice, however, it turns out that the signal acceptance for low-mass LLPs is tiny. The reason is that the DV mass does not necessarily correspond to the mass of the decaying particle. Indeed, the selection requirements introduce a substantial bias of the DV mass distribution to lower values. As a result, even if the LLP has a mass above $10\,\mathrm{GeV}$, a large fraction of events may fail the requirement $m_\mathrm{DV} > 10 \, \mathrm{GeV}$.

\begin{figure}[t]
\centering
\includegraphics[width = 0.6 \textwidth]{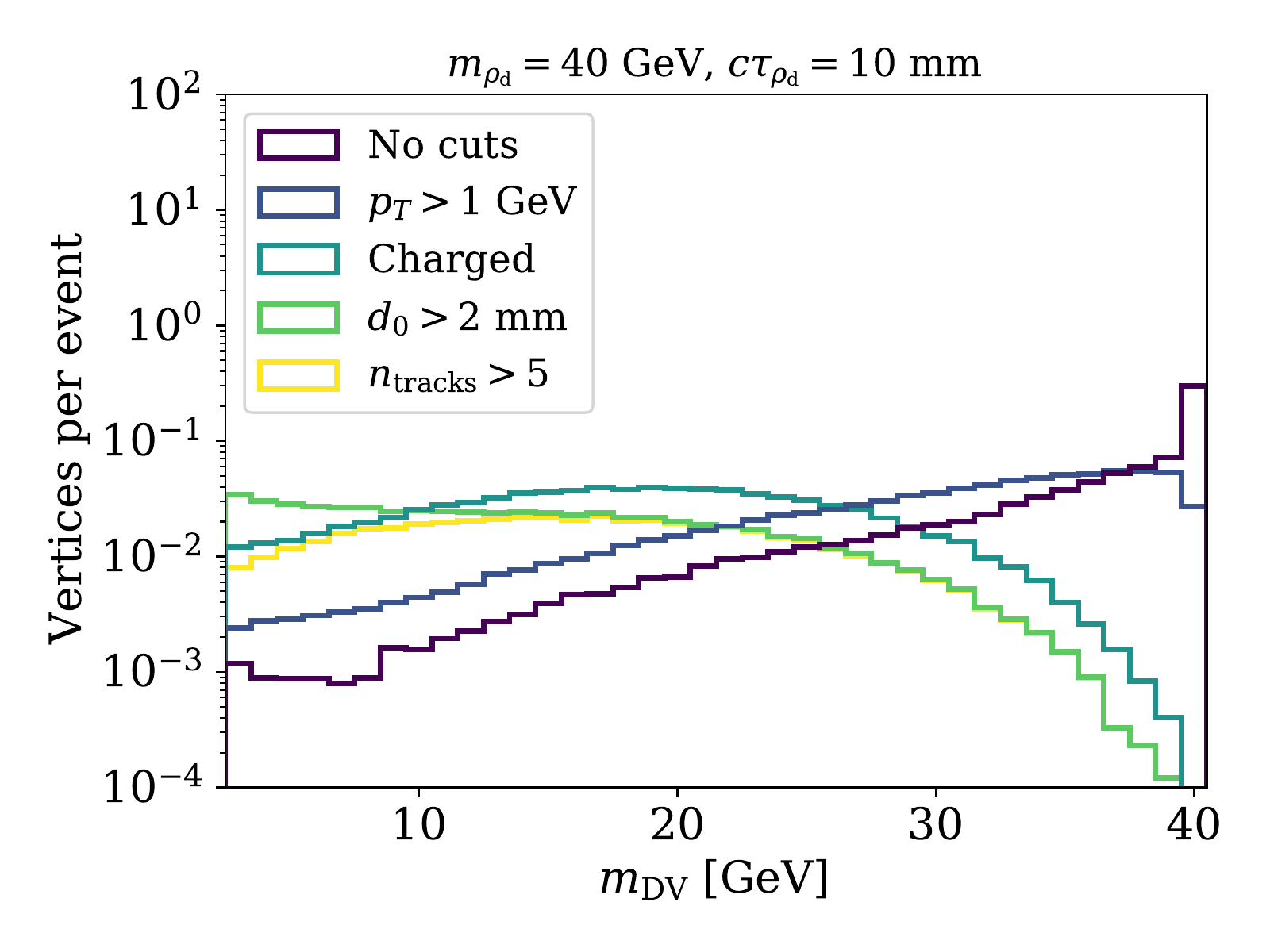}
\caption{Distribution of the mass of vertices reconstructed from decaying dark $\rho^0_\mathrm{d}$ mesons as successive cuts are applied on the particles originating from the vertex. Note that the first two curves include neutral particles, even though these would not be detectable in the tracker.} \label{fig:Mvertex}
\end{figure}

We illustrate this effect in figure~\ref{fig:Mvertex}, which shows the vertex mass distributions after successive cuts for a dark rho meson mass of $\mathrm{40\,\mathrm{GeV}}$ with proper decay length of $10\,\mathrm{mm}$. Cuts are applied in the following order: first the full vertex mass distribution is shown with no requirements ("No cuts") on the tracks, apart from that the vertices fall into the correct position space selection ($R$ and $|z|$) of the search. Next, we show the distribution of vertex mass when tracks forming each vertex are required to have $p_T > 1$~GeV, then that the tracks are charged, then that $d_0 > 2$~mm. Finally, we require the vertex overall to have at least 5 tracks.

When no cuts are applied, the distribution of vertex masses peaks at the true vertex mass $m_{\rho_\mathrm{d}}$. However, the requirement of charged tracks and the $d_0$ cut reduce the number of tracks included in the vertex reconstruction and thereby bias the vertex mass distribution towards smaller values. As a result, only $30\%$ of events with a DV end up satisfying all analysis cuts, significantly affecting the sensitivity of this analysis to GeV-scale LLPs. This effect becomes even more severe for smaller decay lengths, as even more tracks are removed by the $d_0$ requirement.

\subsection{Modified analyses}

To mitigate the bias in the vertex mass introduced by the track requirements, we investigate the effect of (independently) relaxing two classes of cuts:
\begin{itemize}
\item \textit{Relaxed cuts} analysis: Following ref.~\cite{Cottin:2018kmq} we relax the requirements on the DV mass and the number of tracks to $m_{\rm DV} > 5\,\mathrm{GeV}$ and $n_{\rm tracks} \geq 4$.
\item \textit{Relaxed $d_0$} analysis: Motivated by ref.~\cite{ATLAS:2019wqx} we include charged tracks with small impact parameter in the reconstruction of the vertex mass. In order to accurately reconstruct the position of the DV, we still need to require at least two charged tracks with $d_0 > 2\,\mathrm{mm}$.
\end{itemize}

For both modifications, we assume that the signal efficiencies are similar to those from ref.~\cite{Aaboud:2017iio}. Specifically, for the {\it `relaxed cuts'} analysis we assume that the efficiencies are constant below $m_\mathrm{DV} = 10\,\mathrm{GeV}$ and $n_\text{tracks} = 5$. For the {\it `relaxed $d_0$'} analysis, ref.~\cite{ATLAS:2019wqx} only provides efficiencies for three specific models (rather than as a function of experimental observables), so we simply take the efficiency from ref.~\cite{Aaboud:2017iio} for the modified value of $m_\mathrm{DV}$ and the total number of charged tracks.

An accurate estimate of experimental backgrounds for these modified analyses is well beyond the scope of the present work. Nevertheless, we can infer from ref.~\cite{Aaboud:2017iio} that the {\it `relaxed cuts'} analysis would also give 0 observed events based on the current data set. We will therefore assume that the search remains background-free even with the modifications proposed above.

\begin{figure}[t]
	\centering
	
    \includegraphics[height=6.3cm,clip,trim=15 5 95 5]{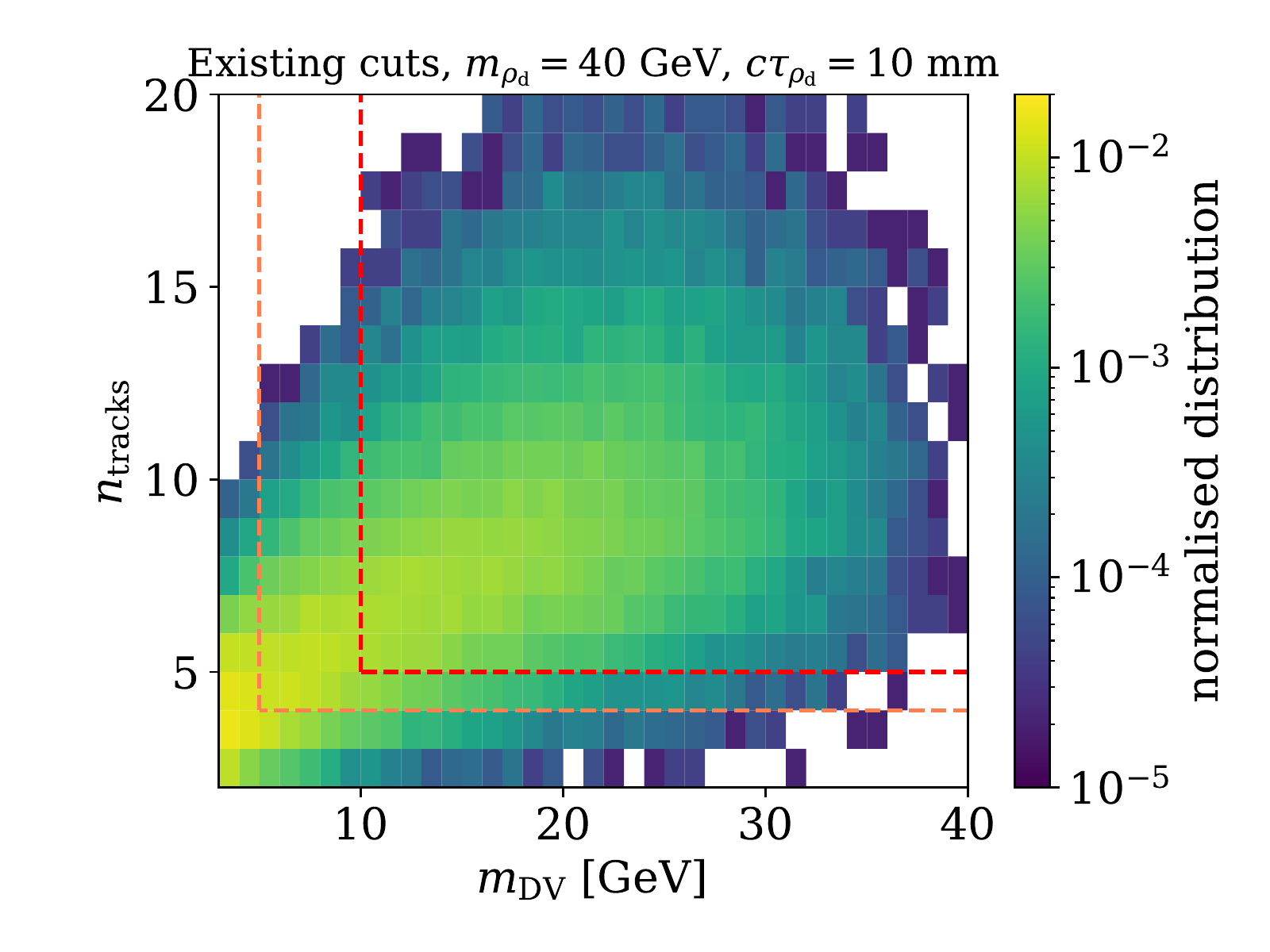}	\includegraphics[height=6.3cm,clip,trim=15 5 15 5]{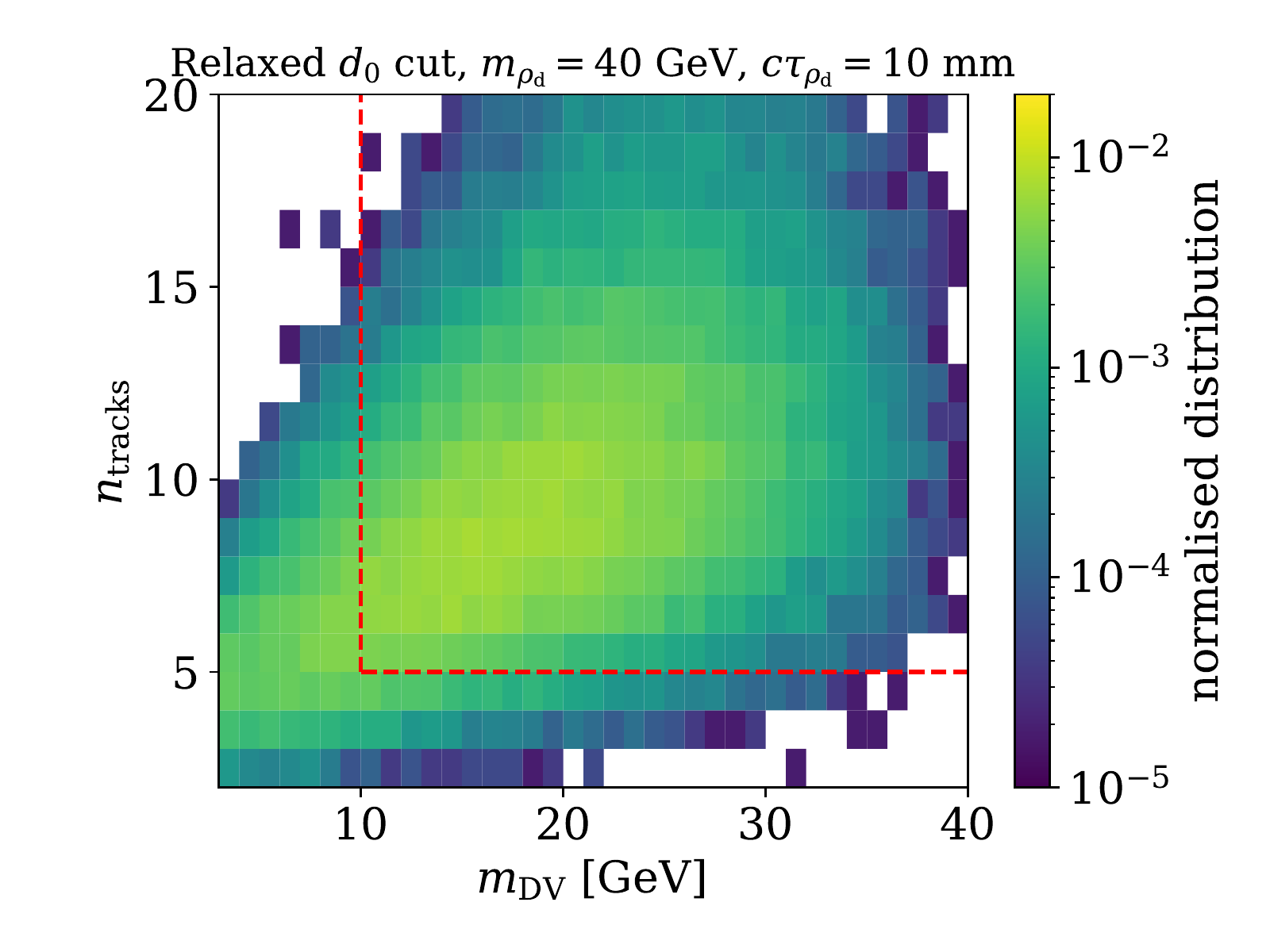}
     \includegraphics[height=6.2cm,clip,trim=15 5 95 5]{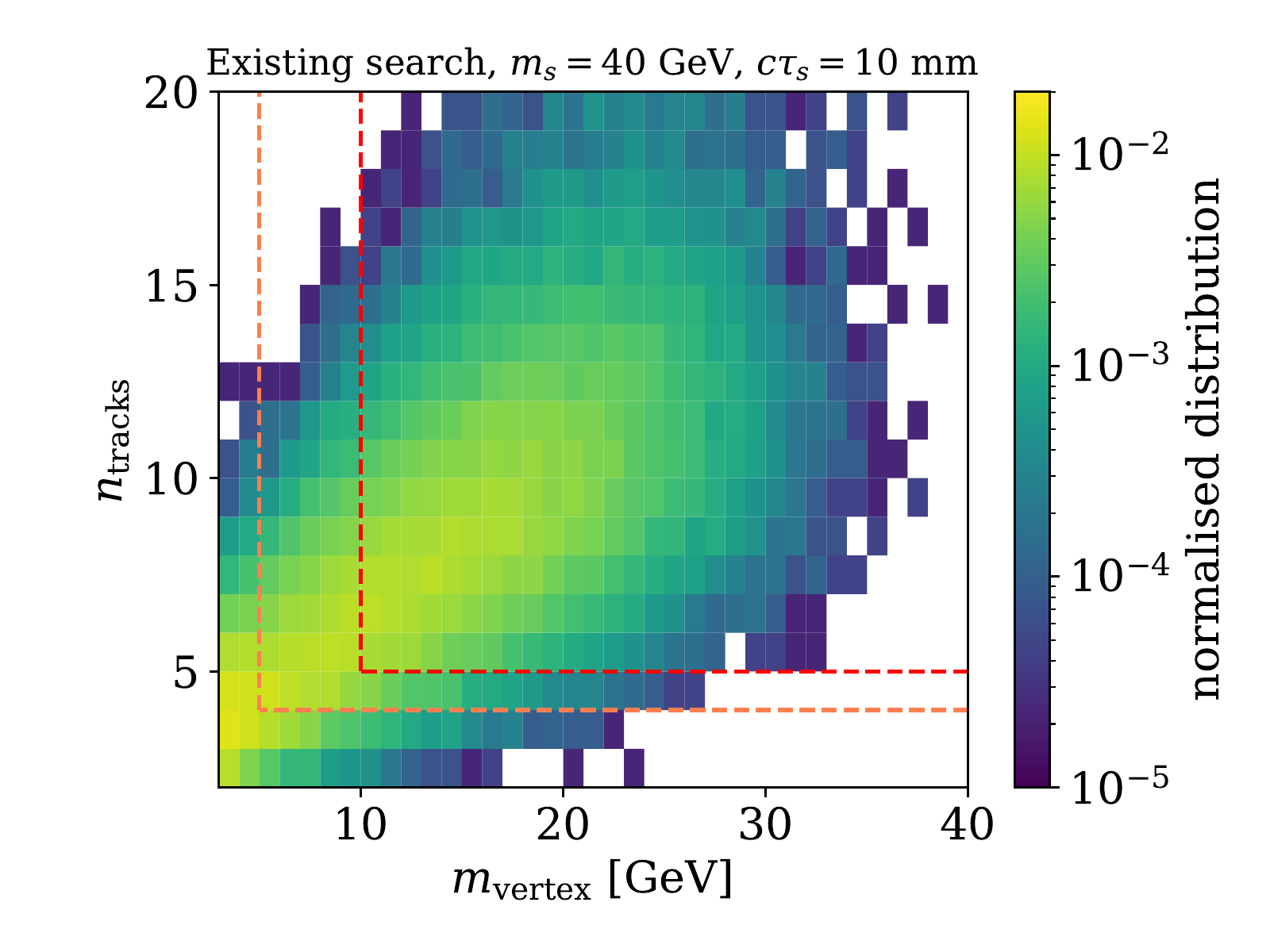}	\includegraphics[height=6.2cm,clip,trim=15 5 15 5]{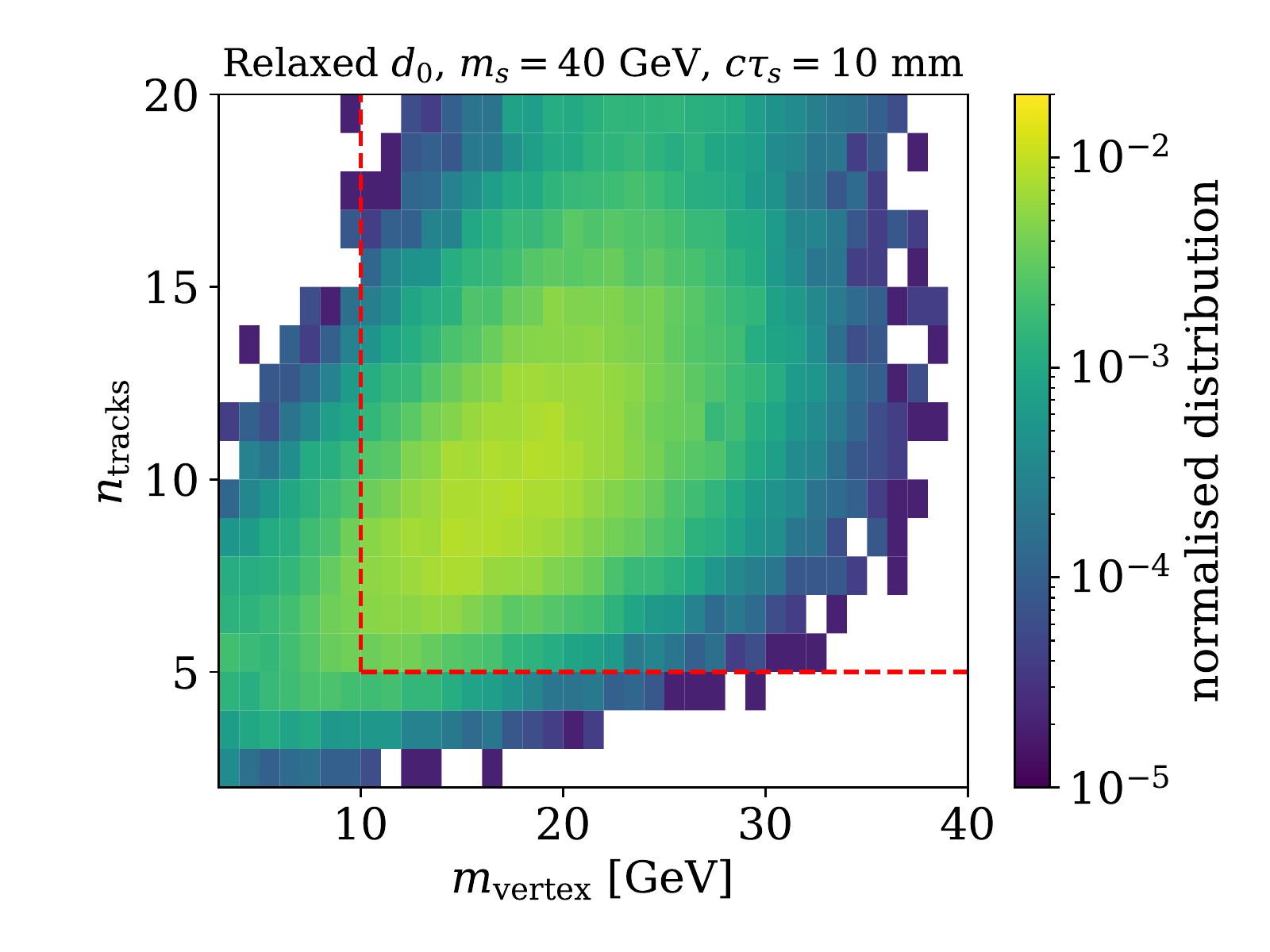}
	\caption{Distribution of vertex mass and number of tracks per displaced vertex originating from dark $\rho^0_\mathrm{d}$ mesons (top) and dark Higgs bosons (bottom) after the track-level cuts of the existing analysis (left) and of the {\it `relaxed $d_0$'} modification (right). The search window is shown with dashed lines. On the left, we additionally show the search window of the {\it `relaxed cuts'} modification.\label{fig:mvertex_ntrack}}
\end{figure}

%

We illustrate the proposed modifications in figure~\ref{fig:mvertex_ntrack}, which shows the two-dimensional distribution of vertex mass and number of tracks per vertex for the existing analysis cuts (left) and the relaxed $d_0$ analysis (right). Again, we consider the strongly-interacting dark sector model for concreteness. We indicate the search window for the existing and the {\it `relaxed $d_0$'} analysis with red dashed lines in both panels, while the search window for the {\it `relaxed cuts'} analysis is indicated with an orange dashed line in the left panel. With the cuts of the existing analysis, the bulk of the distribution for a low-mass LLP lies outside the search window. The {\it `relaxed $d_0$'} modification shifts the distribution to larger $n_\mathrm{track}$ and $m_\mathrm{DV}$ and thus into the search window. The {\it `relaxed cuts'} analysis, on the other hand, expands this window to encompass most of the distribution without modifying other track requirements.

\section{Results}
\label{sec:results}

Having selected all displaced vertices that fall into the search window, we can now apply the vertex-level and event-level efficiencies in order to calculate the total efficiency for the existing search and each proposed modification. This efficiency can be interpreted as the expected number of displaced vertices within the search window per event. Note that for the case of a strongly interacting dark sector there can be several dark rho mesons per event and hence the efficiency can~-- at least in principle~-- be larger than unity.

\begin{figure}[t]
\centering

\includegraphics[width = 0.495 \textwidth]{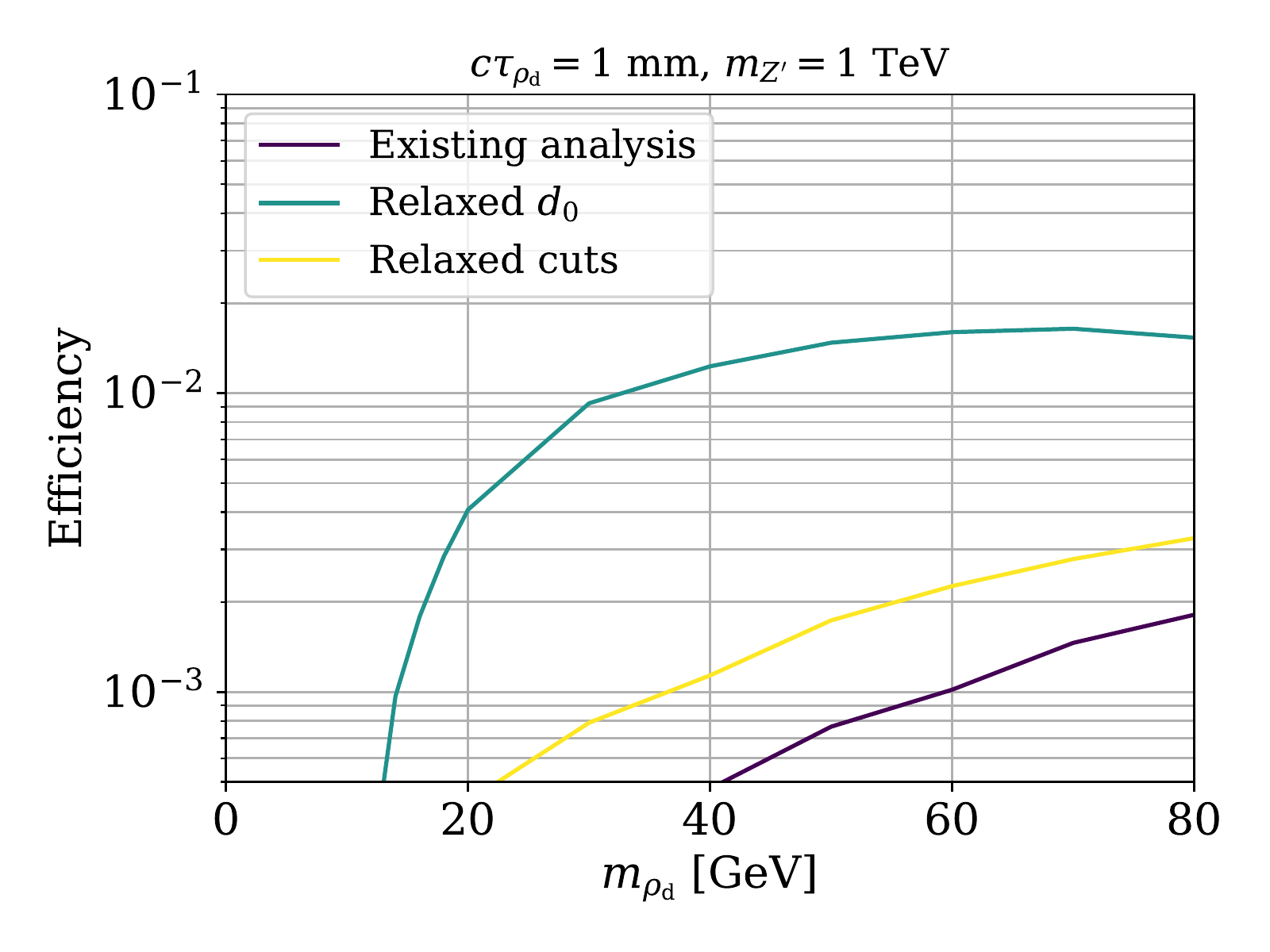}
\includegraphics[width = 0.495 \textwidth]{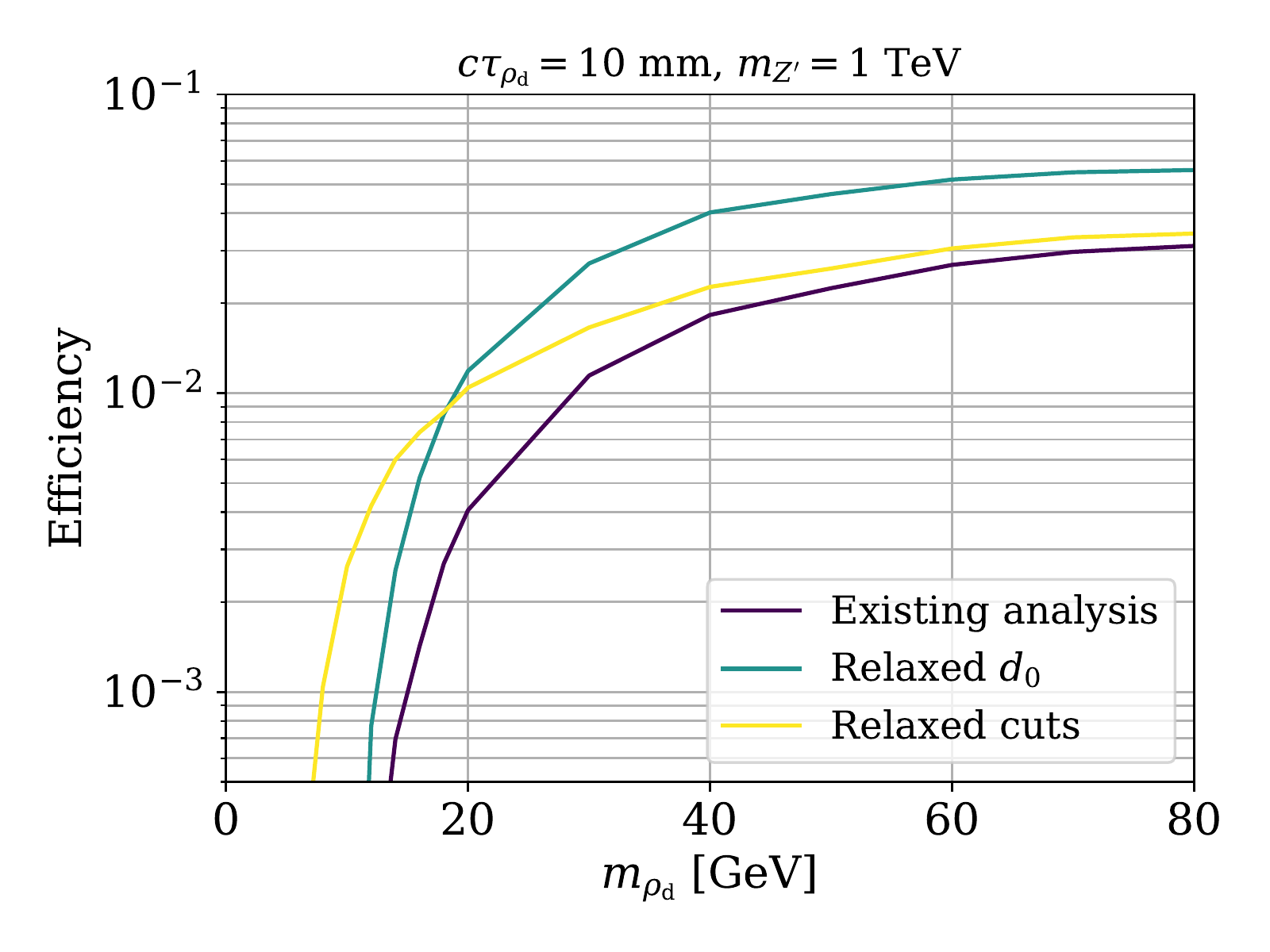}

\includegraphics[width = 0.495 \textwidth]{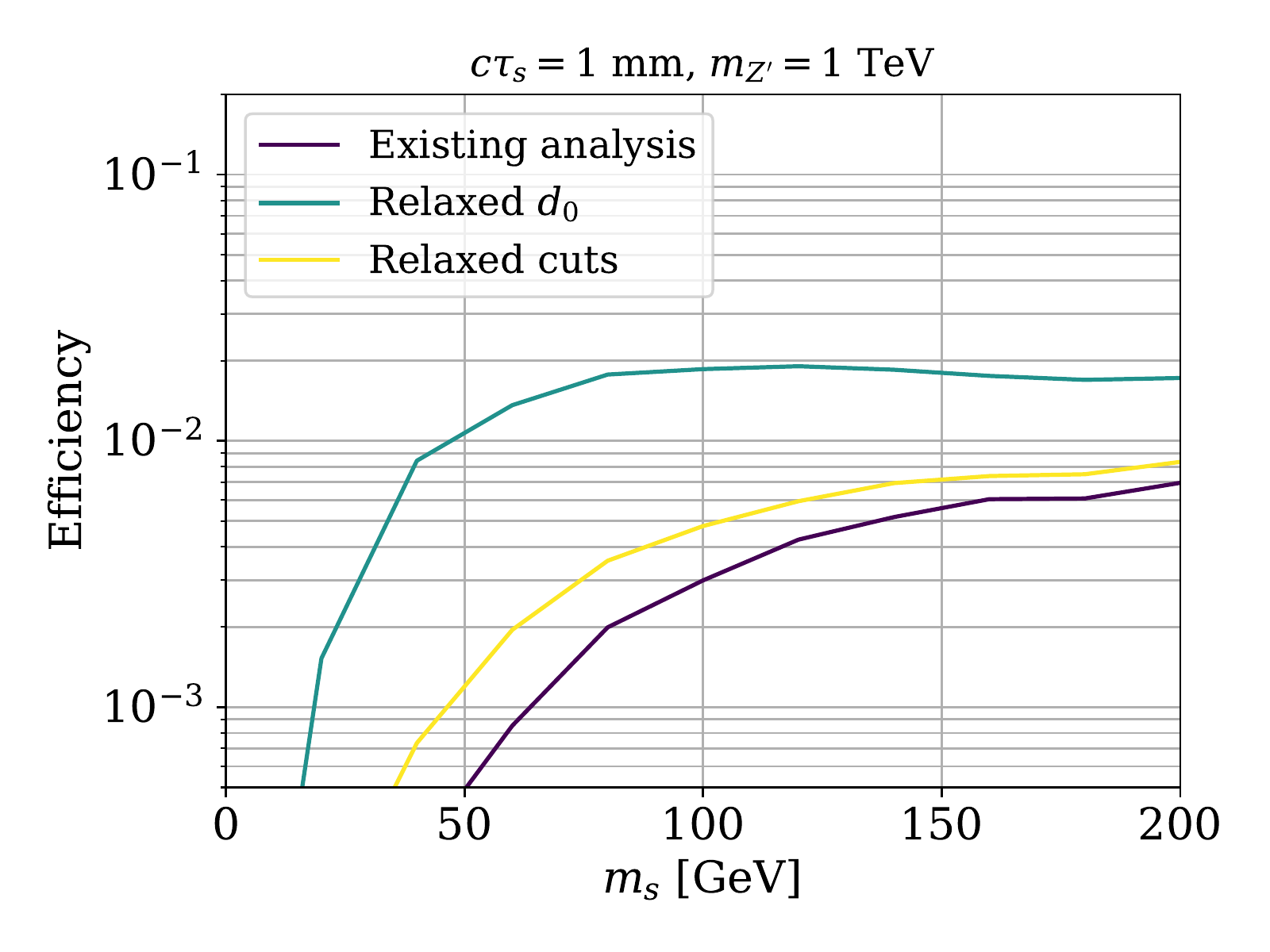}
\includegraphics[width = 0.495 \textwidth]{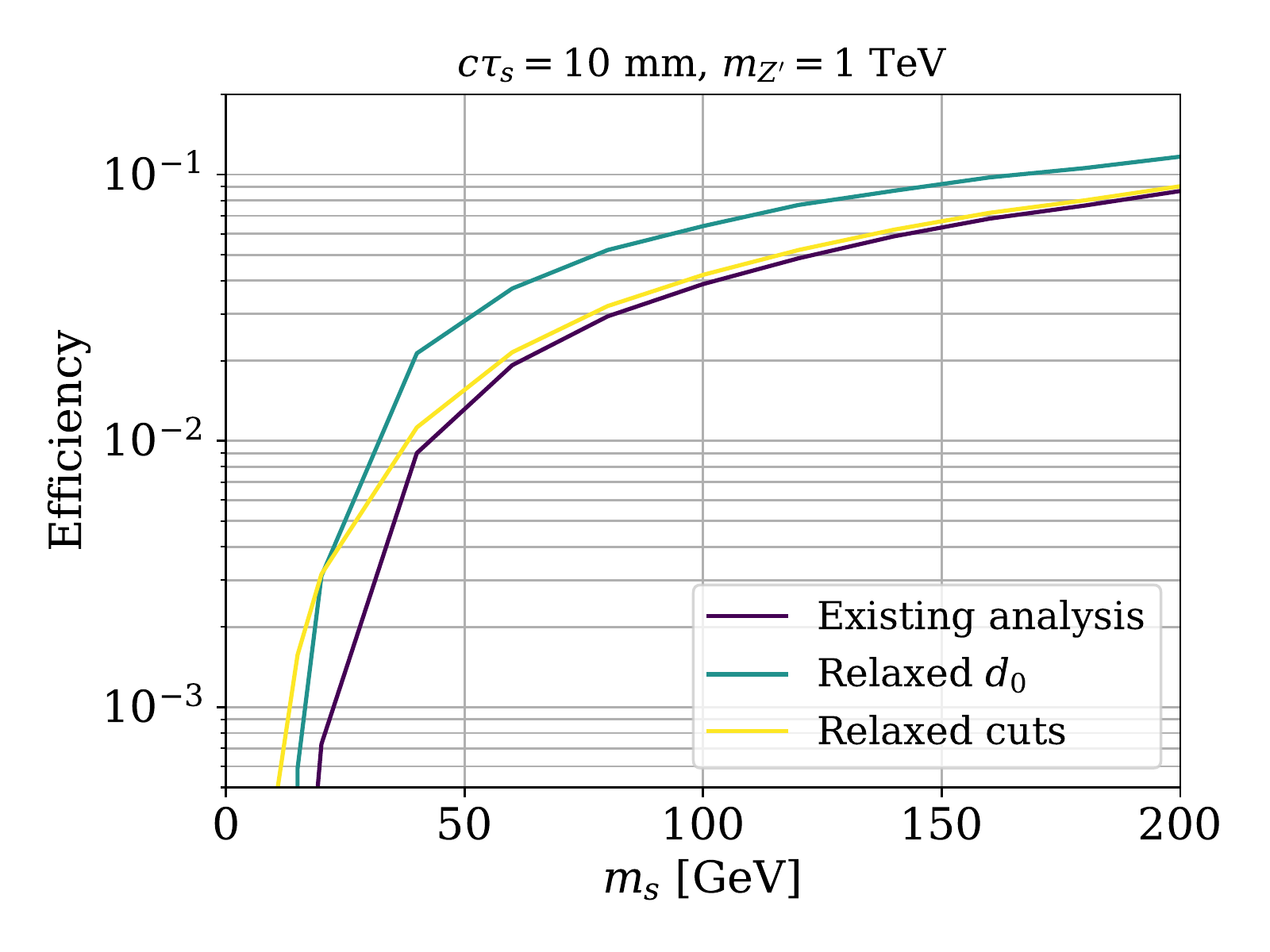}

\caption{Efficiencies for the dark shower model (top) and the dark Higgs model (bottom) as a function of the LLP mass for the existing analysis, compared to the two proposed relaxations. The LLP lifetime is fixed to 1~mm (left) and 10~mm (right).} \label{fig:Effs}
\end{figure}

We show this efficiency for two fixed lifetimes as a function of the LLP mass $m_{\rho_\mathrm{d}}$ of our dark shower signal in the top row of figure~\ref{fig:Effs}.
We observe that the efficiency rises steeply with increasing $m_{\rho_\mathrm{d}}$ even for LLP masses well above the nominal vertex mass cut and only levels off at LLP masses much larger than $10$~GeV. This is in accordance with figure~\ref{fig:mvertex_ntrack}, where we found the distribution of $m_\text{   vertex}$ to peak at values much smaller than $m_{\rho_\mathrm{d}}$. Compared to the existing and the {\it`relaxed cuts'} analysis, the {\it`relaxed $d_0$'} analysis has a distinct advantage at small LLP lifetime, where the majority of tracks from LLP decays have impact parameters below $2$~mm. For instance, at $c\tau_{\rho_\mathrm{d}}=1$~mm and $m_{\rho_\mathrm{d}}=40\,\mathrm{GeV}$ the {\it`relaxed $d_0$'} analysis surpasses the {\it`relaxed cuts'} analysis in efficiency by an order of magnitude and the existing analysis by a factor 20. For larger lifetimes and small masses, on the other hand, the {\it `relaxed cuts'} analysis shows the highest signal efficiency among the proposed modifications. The corresponding efficiencies for the dark Higgs signal are shown in the bottom row of figure~\ref{fig:Effs}. They exhibit qualitatively similar behaviour, but rise even more slowly with increasing LLP mass.

Up to now we have focused exclusively on the efficiency and have not accounted for the dependence of the signal cross section on the LLP mass. While figure~\ref{fig:Effs} shows that the efficiencies at fixed LLP lifetime generally increase as a function of the LLP mass, the cross section for dark shower production falls with increasing $m_{\rho_\mathrm{d}}$ if we hold the lifetime fixed. We emphasise that this is not a kinematic effect. Since the lifetime scales like
\begin{align}
c\tau_{\rho_\mathrm{d}} \propto  g_q^{-2} \, e_\mathrm{d}^{-2} \, m_{\rho_\mathrm{d}}^{-5} \; ,
\end{align}
the couplings need to decrease in order to keep the lifetime constant when increasing the dark rho mass. Smaller couplings in turn reduce the dark shower production cross section, which scales with $g_q$ and the $Z'$ branching ratio to dark quarks as
\begin{align}
\sigma_{p p \to q_\mathrm{d} \bar{q}_\mathrm{d}} \propto  g_q^2 \, \mathrm{BR}\left(Z' \to q_\mathrm{d} \bar{q}_\mathrm{d}\right) = g_q^2 \, \frac{1}{1 + 3\left(\frac{g_q}{e_d}\right)^2} \; .
\end{align}
Hence, if we hold $c\tau_{\rho_\mathrm{d}}$ and the coupling ratio $g_q/e_\mathrm{d}$ fixed,\footnote{This is equivalent to varying only the underlying $U(1)'$ gauge coupling but not the (dark) quark charges.} we find that
$\sigma_{p p \to q_\mathrm{d} \bar{q}_\mathrm{d}} \propto  m_{\rho_\mathrm{d}} ^{-5/2}$.

\begin{figure}[t]
\centering
\includegraphics[width=0.495\columnwidth]{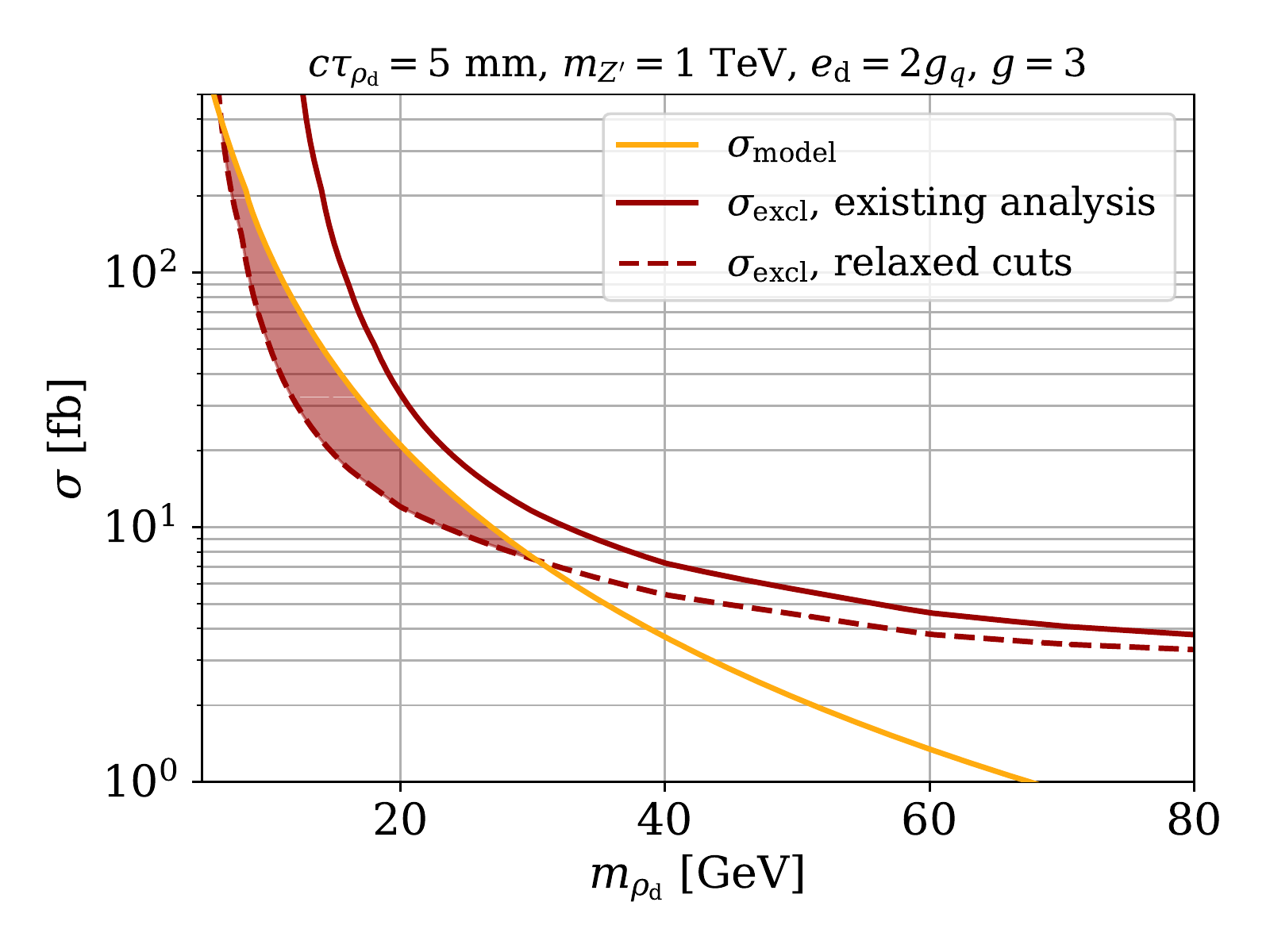}
\includegraphics[width=0.495\columnwidth]{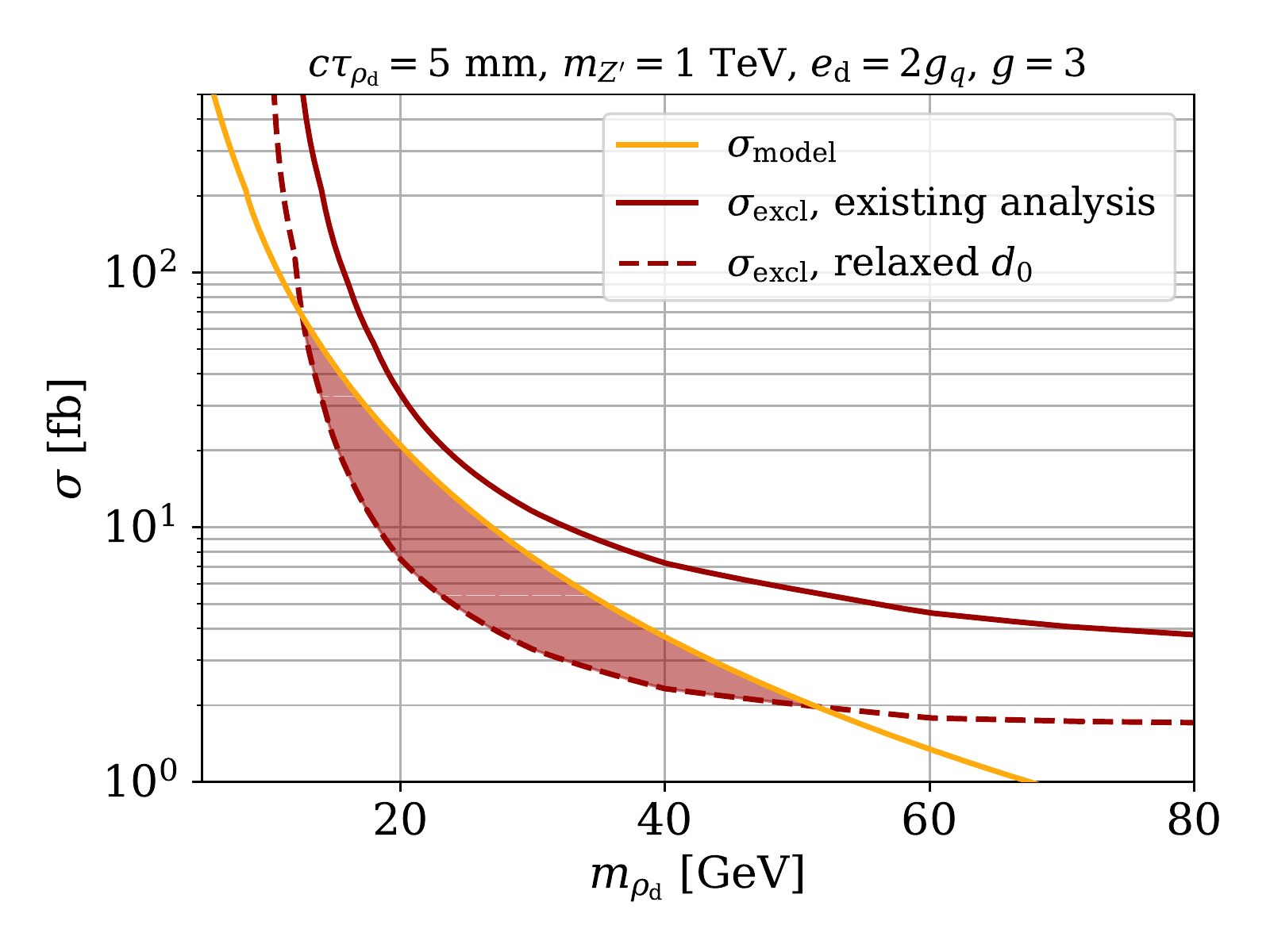}

\includegraphics[width=0.495\columnwidth]{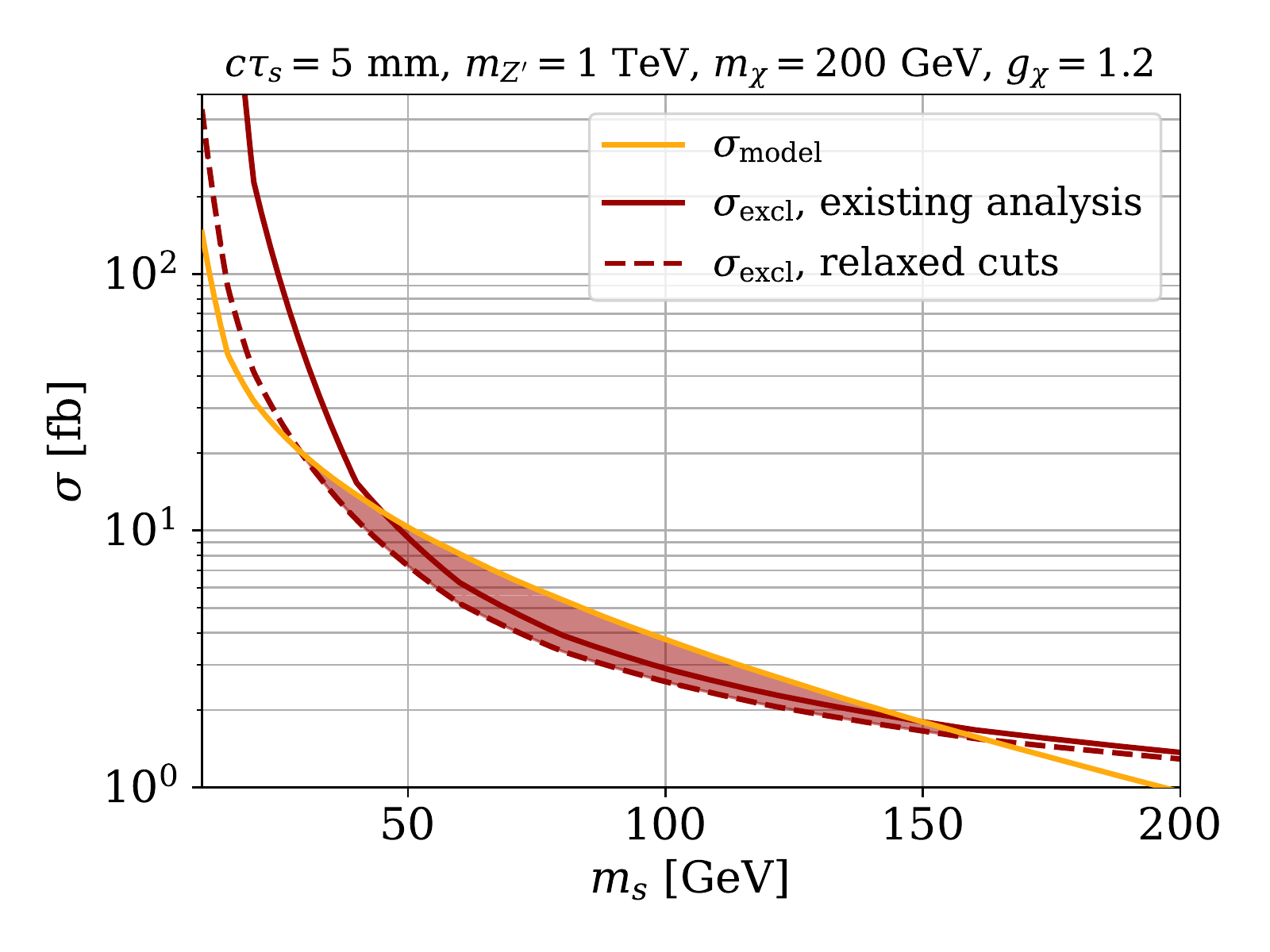}
\includegraphics[width=0.495\columnwidth]{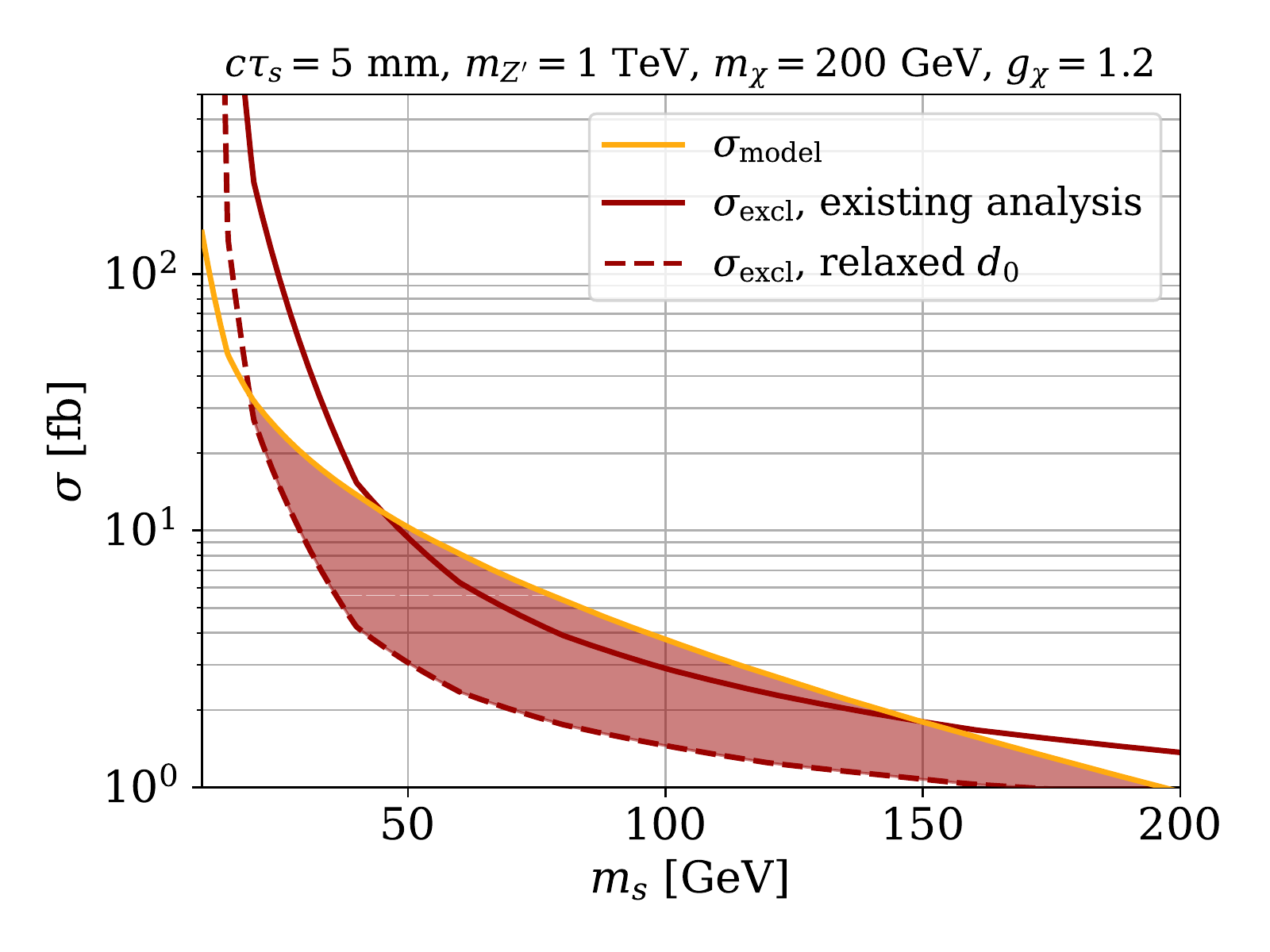}
\caption{Dark shower (top) and dark Higgs (bottom) production cross section in our model compared to the exclusion limit on the cross section from the existing search as well as the {\it `relaxed cuts'} analysis (left) and the {\it `relaxed $d_0$'} analysis (right) at $32.8$~fb$^{-1}$. The LLP lifetime is set to a fixed value, leading to an indirect dependence of the model cross section on the LLP mass. \label{fig:crosssection_excl}} 
\end{figure}

This indirect dependence of the dark shower production cross section on the dark rho mass for fixed lifetime is illustrated by the falling orange line in the top row of figure~\ref{fig:crosssection_excl}. To derive (expected) bounds on the model parameter space, this model cross section needs to be compared to the 95\% confidence limit $\sigma_\mathrm{excl}$ on the cross section obtained from the existing search and our two proposed modifications. Since we assume the search to be free from backgrounds even with the proposed modifications, we can exclude parameter points that predict at least 3 displaced vertices passing all selection requirements. Hence, the excluded cross section is simply $\sigma_\mathrm{excl} = 3/(\mathcal{L}\epsilon)$, where $\epsilon$ denotes the total efficiency discussed above. In figure~\ref{fig:crosssection_excl} we show $\sigma_\mathrm{excl}$ for the existing analysis, the {\it `relaxed $d_0$} analysis and the {\it `relaxed cuts'} analysis as a function of $m_{\rho_\mathrm{d}}$ for fixed $c\tau_{\rho_\mathrm{d}}=5$~mm, $g_q/e_\mathrm{d}=0.5$ and $m_{Z'}=1$~TeV and an integrated luminosity of 32.8~fb$^{-1}$. For these particular parameters the existing analysis yields an exclusion limit that comes close to the model prediction, but does not make an exclusion, while both modifications can probe the predicted cross section. A red shading indicates the range of dark rho masses to which the modified analyses are sensitive. As expected, the {\it `relaxed cuts'} modification can reach to smaller masses, while the {\it `relaxed $d_0$'} analysis is more sensitive at larger masses of up to 50~GeV for the particular model parameters chosen here. 

The corresponding plots for the dark Higgs model are shown in the bottom row of figure~\ref{fig:crosssection_excl}. In contrast to the dark rho meson case, the existing search does make a modest exclusion for a luminosity of $32.8\,\mathrm{fb^{-1}}$, which is only slightly extended with the {\it `relaxed cuts'} analysis. For the {\it `relaxed $d_0$'} analysis we find a much larger region of sensitivity, which however lies at significantly larger masses than for the dark rho mesons. The increased sensitivity towards larger masses is a result of the much weaker mass dependence of the production cross section for fixed lifetime. Since $c\tau_{s} \propto g_q^4 g_\chi^2 m_s$ we find that for fixed $c\tau_{s}$ and fixed $g_\chi$ the production cross section scales as $\sigma_{p p \to s + X} \propto g_q^2 \propto m_s^{-1/2}$.

\begin{figure}[t]
	\centering
	\includegraphics[width=0.495\columnwidth]{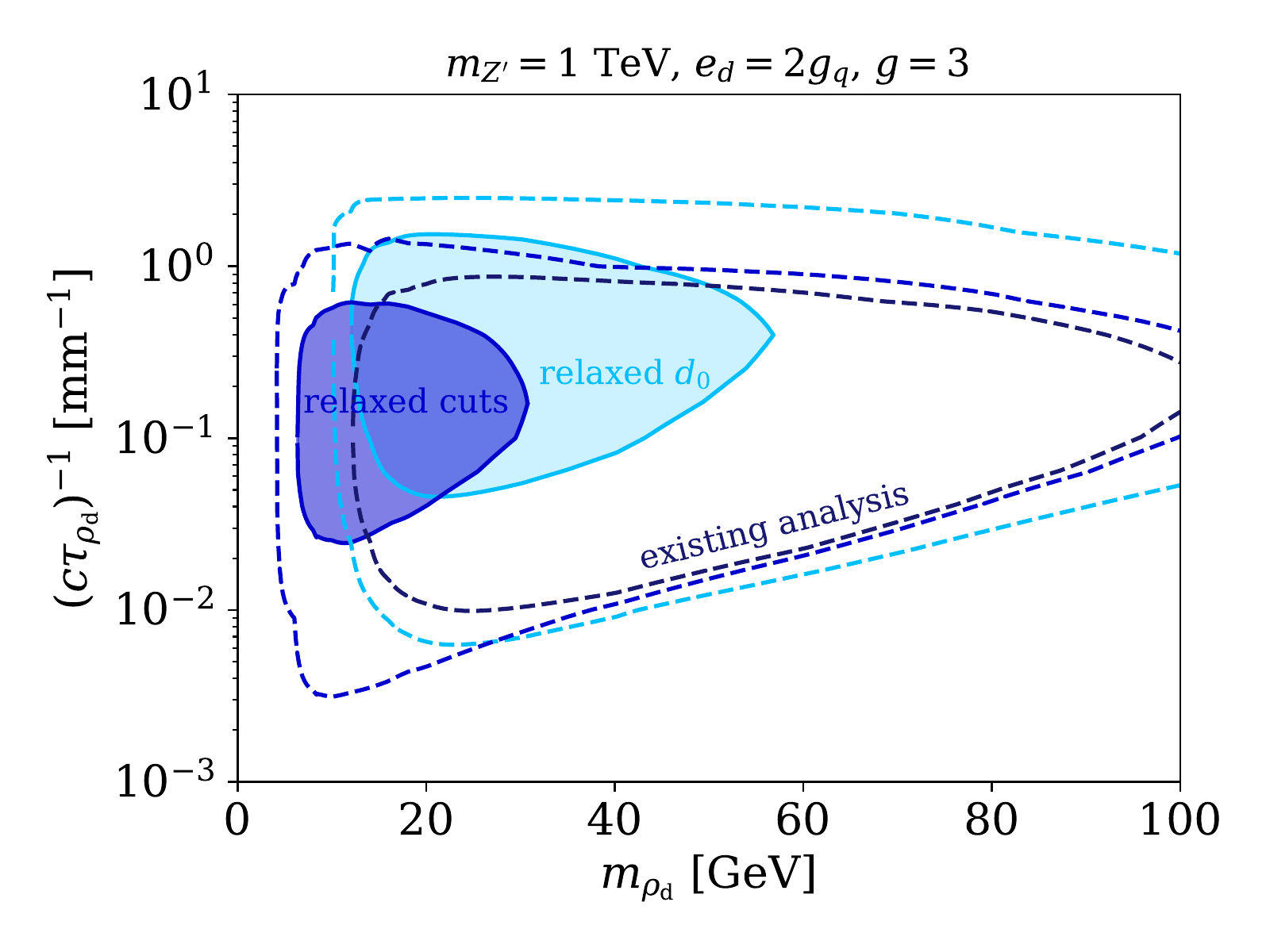}
	\includegraphics[width=0.495\columnwidth]{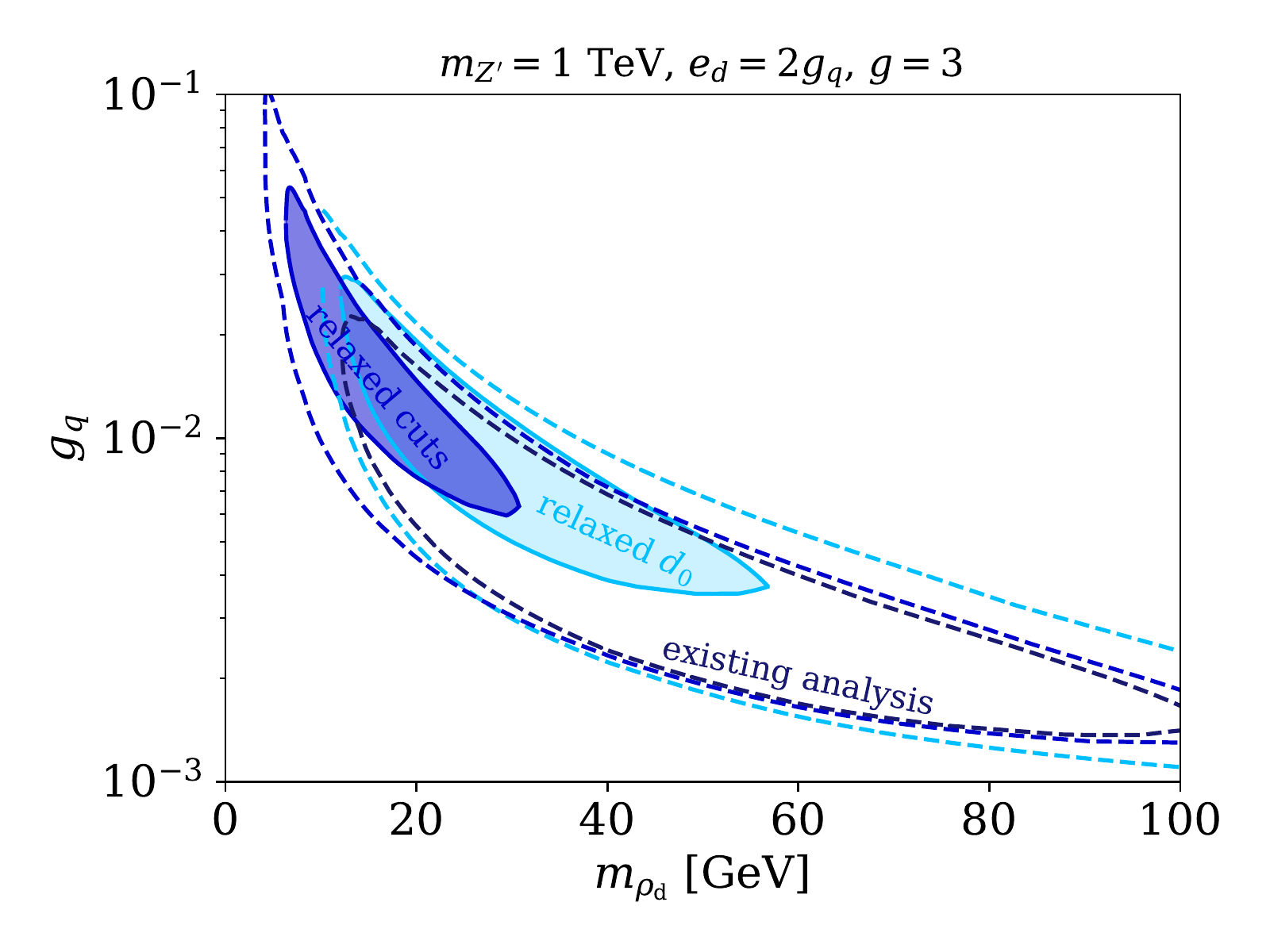}

	\includegraphics[width=0.495\columnwidth]{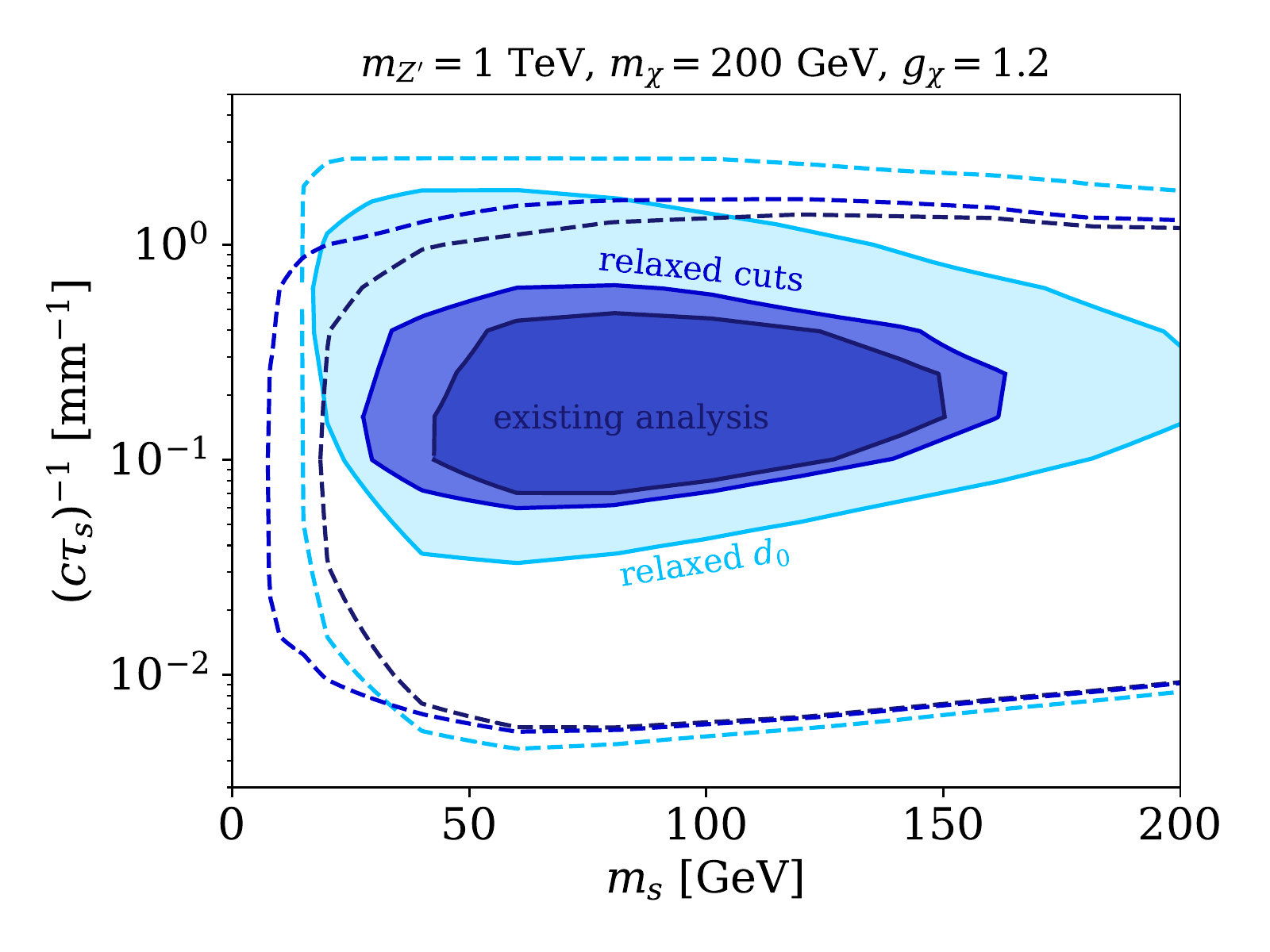}
	\includegraphics[width=0.495\columnwidth]{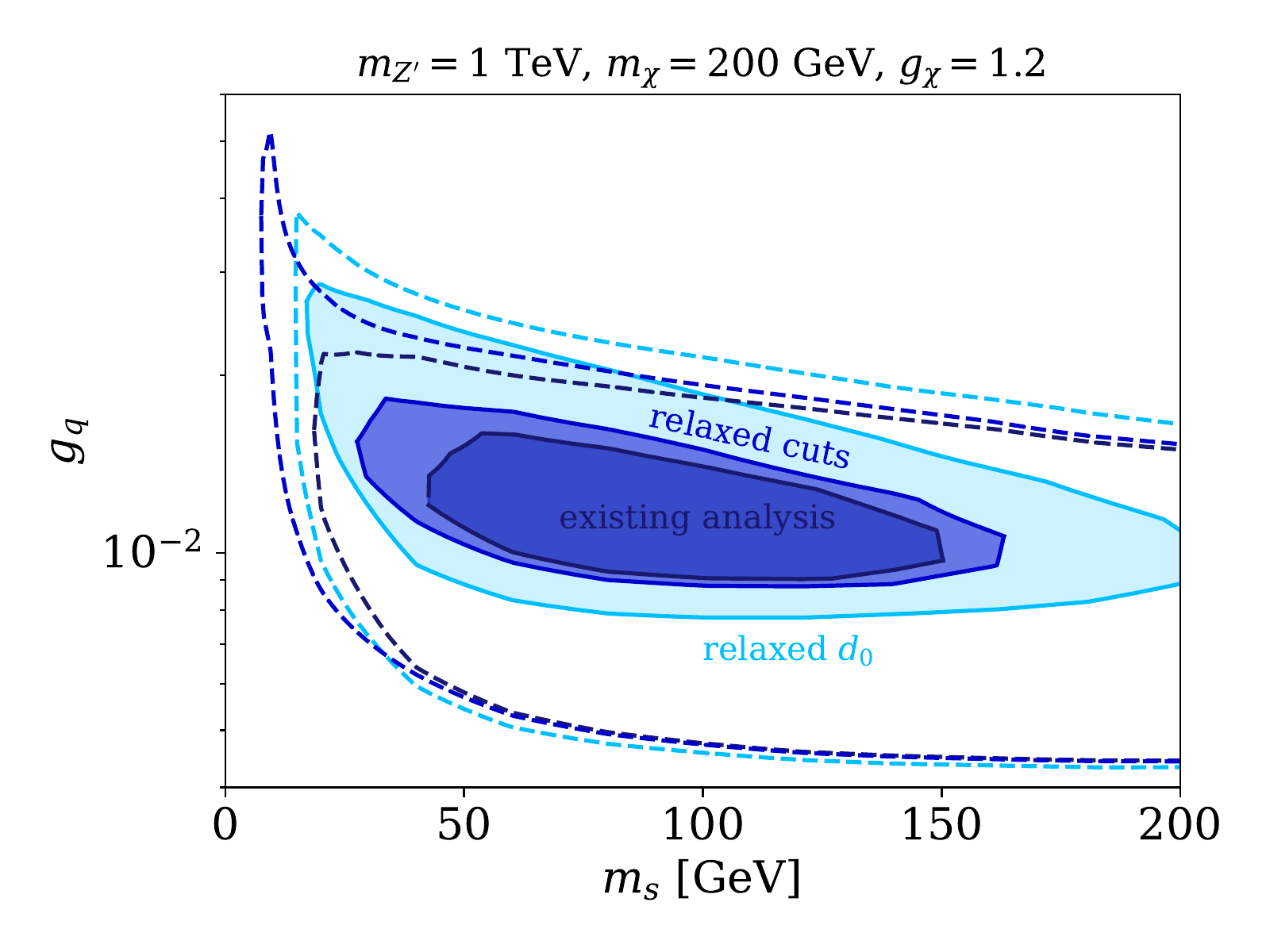}
	
	\caption{(Projected) exclusion contours of the existing analysis, the {\it `relaxed $d_0$'} analysis, and the {\it `relaxed cuts'} analysis for the dark shower model (top) and the dark Higgs model (bottom) shown in the parameter plane spanned by the LLP mass and its inverse lifetime (left) or coupling $g_q$ (right). Contours for 32.8~fb$^{-1}$ integrated luminosity are shown with solid lines, contours for 300~fb$^{-1}$ with dashed lines. \label{fig:dvmet_excl}}
\end{figure}

In the next step we also scan over $\tau_{\rho_\mathrm{d}}$ and determine an upper and a lower (expected) bound on $m_{\rho_\mathrm{d}}$ for each lifetime and each analysis. As a result we obtain the two-dimensional sensitivity contours displayed in the top-left panel of figure~\ref{fig:dvmet_excl}. Again the existing analysis does not make an exclusion at 32.8~fb$^{-1}$, so we only show the projections for our two proposed modifications. As before, the {\it `relaxed cuts'} analysis is sensitive to lower masses. More interestingly, it also becomes obvious in this plot that the {\it `relaxed $d_0$'} analysis is sensitive to smaller lifetimes or shorter decay lengths, even below 1~mm. As we go to larger $c\tau_{\rho_\mathrm{d}}$, the two analyses lose sensitivity at decay lengths of approximately 20~mm and 50~mm, respectively, owing to the decreasing cross section at longer decay lengths which correspond to smaller coupling. This coupling $g_q$ is shown instead of $c\tau_{\rho_\mathrm{d}}$ on the vertical axis of the top-right panel of figure~\ref{fig:dvmet_excl} and spans over a range between $g_q=0.003$ and $gq=0.05$ in the excluded contours of the two analyses.

The projections discussed so far assumed the same luminosity as in the existing search. In addition, we also include in figure~\ref{fig:dvmet_excl} projections for 300~fb$^{-1}$ (shown in dashed lines) to account for the increase in integrated luminosity expected after LHC run 3. To calculate these projections we assume that the search remains background free even with the larger data set (see below for further discussion). Higher luminosity greatly expands the reach of the search towards larger masses and longer decay lengths. The gains towards shorter decay lengths are more modest as the displaced vertex cuts require the average LLP decay length to be not much shorter than a millimetre.

The corresponding plots for dark Higgs decays are shown in the bottom row of figure~\ref{fig:dvmet_excl}. Since the dark Higgs lifetime depends sensitively on the coupling strength $g_q$, the {\it `relaxed cuts'} and {\it `relaxed $d_0$'} analyses turn out to be sensitive to a relatively narrow range of couplings around $g_q \approx 0.01$ (see bottom-right panel of figure~\ref{fig:dvmet_excl}). Nevertheless, with increasing luminosity this range can be extended considerably to both larger and smaller couplings. The range of dark Higgs masses that can be probed also increases substantially. Note that we do not consider $m_s > 200\,\mathrm{GeV}$, because the model we consider requires the hierarchy $m_\chi > m_s$. However, for different values of $m_\chi$ (or relaxing the model assumptions), we expect that the search should be sensitive also to heavier dark Higgs bosons.

Let us finally comment on complementary search strategies. For the case of a strongly-interacting dark sector, the dark rho meson can mediate interactions between the dark pions and SM particles. If the dark pions are stable and we fix their mass through the assumption that they constitute all of DM, we hence obtain complementary bounds from direct detection experiments. These bounds are found to be sensitive to a comparable range of couplings, excluding $g_q \gtrsim 0.01$ for $m_{\rho_\mathrm{d}} \gtrsim 20\,\mathrm{GeV}$. However, since these bounds require additional assumptions, we do not show them in figure~\ref{fig:dvmet_excl}. For the dark Higgs model, bounds from direct detection are suppressed due to the assumed Majorana nature of the DM particle and only probe $g_q \gtrsim 0.5$. The strongest complementary constraints hence arise from LHC searches for jets in association with missing energy, which exclude $g_q \gtrsim 0.1$. We are not aware of any existing constraints sensitive to the coupling range considered in figure~\ref{fig:dvmet_excl}, emphasizing the unique power of LHC searches for long-lived particles.\footnote{We note that ATLAS has recently carried out a search for dark Higgs bosons decaying into $W^+W^-$ and $ZZ$~\cite{Aad:2020sef}. However, these decay channels are suppressed in the case that the dark Higgs decays via loop-induced processes rather than through mixing with the SM Higgs. Hence, this search is not sensitive to the scenario considered here.}

\bigskip

To conclude this section, we comment on the case of non-negligible backgrounds. Indeed, it is likely that the searches proposed here will not remain background-free  with increasing luminosity. This is particularly true when relaxing the cut on $d_0$, which is introduced to reduce backgrounds from tracks originating from the interaction point accidentally crossing a DV.

First of all, we emphasize that the searches that we consider  may be optimised further to suppress backgrounds while simultaneously maintaining sensitivity to dark sectors. For example, relaxed DV requirements may be compensated by stricter cuts on the overall event topology, for example on the amount of missing energy. Moreover, backgrounds typically decrease with increasing transverse distance from the interaction point, such that a background-free search may be possible at least for LLPs with comparably long decay length.

In the case that non-negligible backgrounds remain, it will likely be challenging to obtain an accurate background model, making standard methods for background subtraction unsuitable. Nevertheless, there are various methods that can be used to set limits in the presence of an unknown background. The simplest such method is to treat all observed events as signal and calculate a Poisson upper limit. The generalisation to several signal regions known as the ``Binned Poisson'' method~\cite{Savage:2008er} is frequently used by direct detection experiments facing unknown backgrounds. Unbinned methods also exist, such as the maximum gap or optimum interval method~\cite{Yellin:2002xd}.

\begin{figure}[t]	
	\includegraphics[width=0.495\columnwidth]{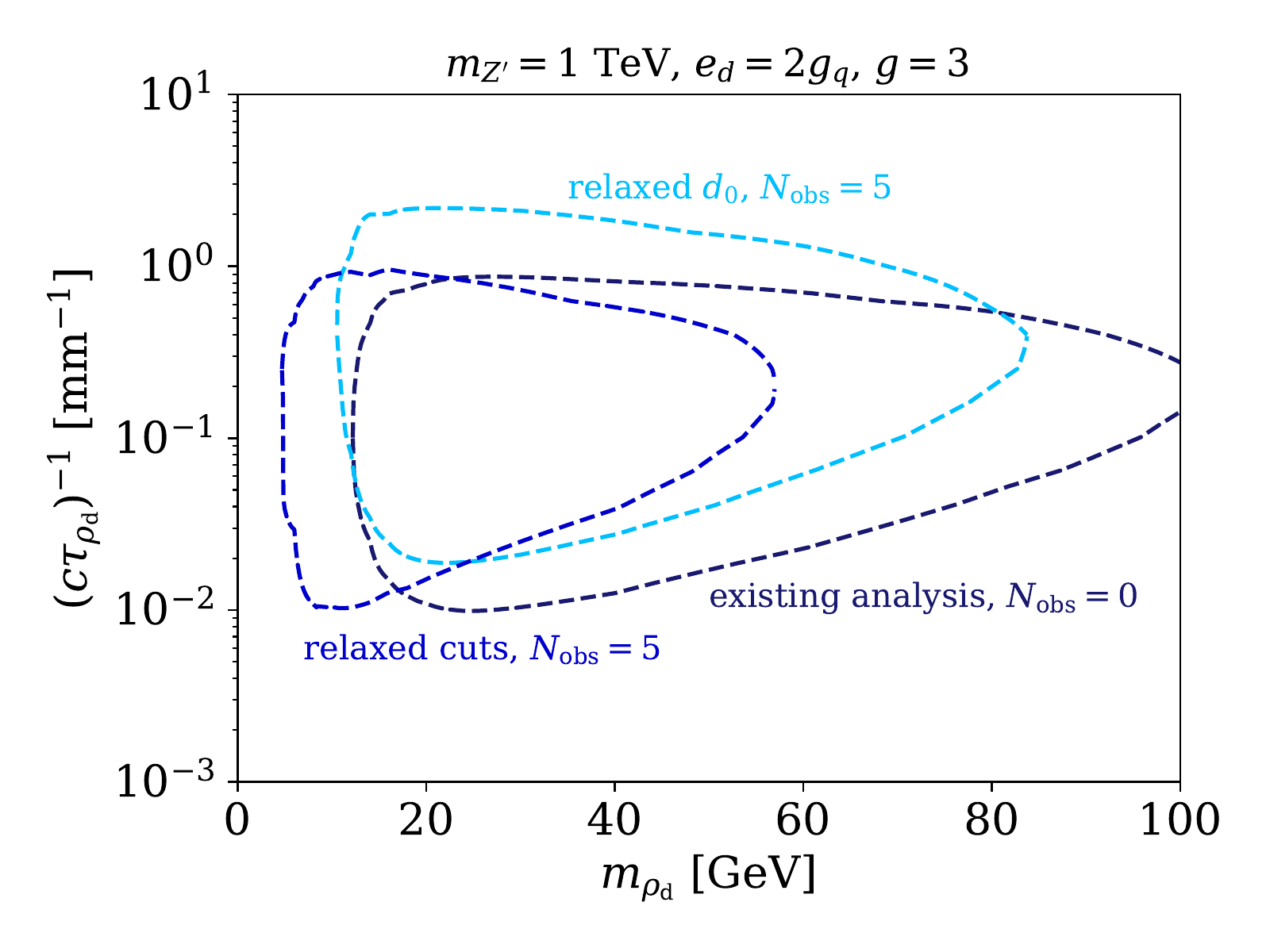}
	\includegraphics[width=0.495\columnwidth]{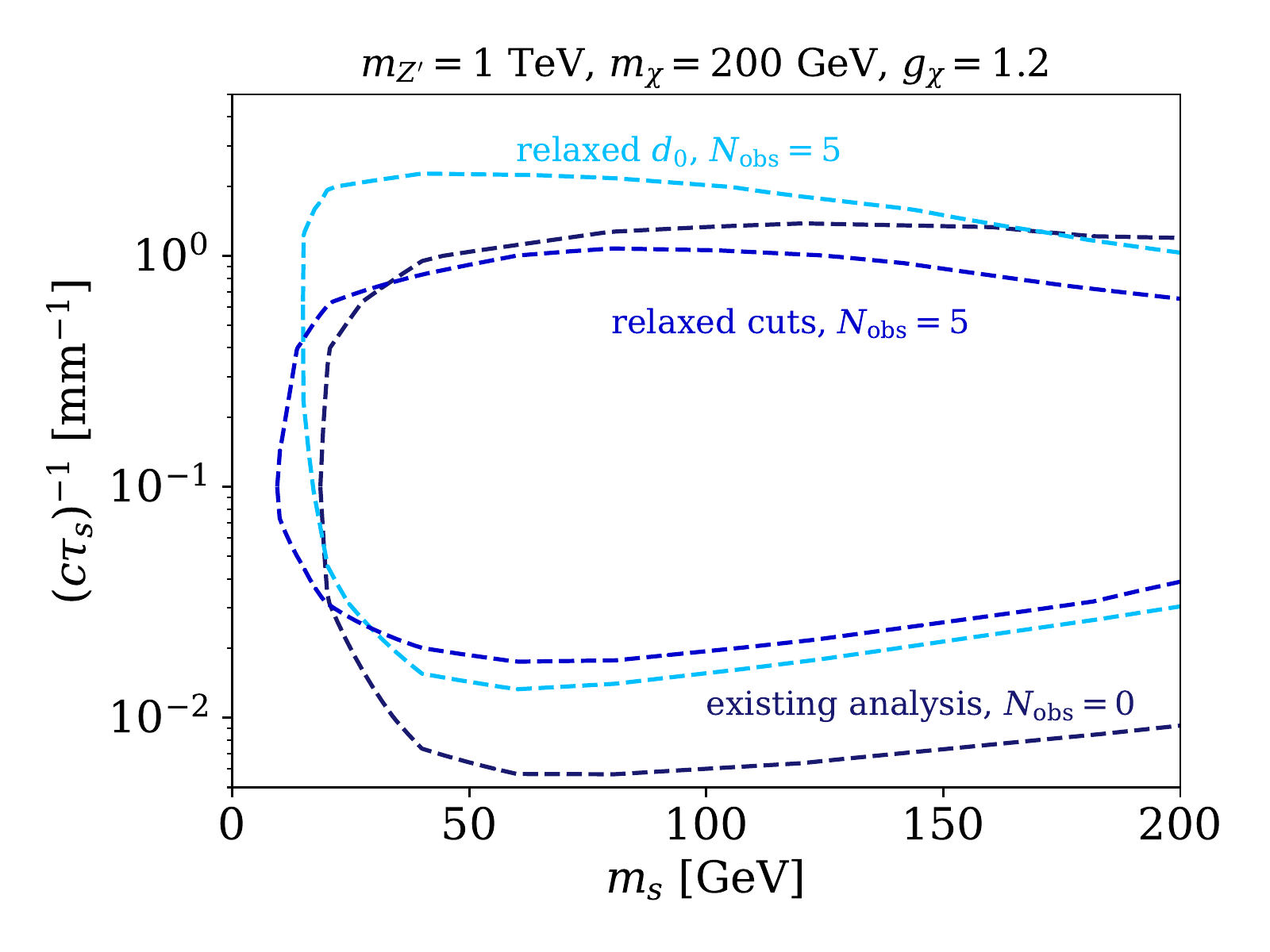}
	
	\caption{Projected exclusion contours for 300~fb$^{-1}$ for the dark shower model (left) and the dark Higgs model (right) under the assumption that each of the modified analyses leads to 5 observed events, while the original analysis remains background-free. Note that we conservatively assume that no background subtraction is performed (see text for details).\label{fig:dvmet_excl_nobs5}}
\end{figure}

In the case of a single signal region, an observation of 1 (5) events would allow one to exclude any signal hypothesis predicting more than 4.7 (10.5) events at 95\% confidence level without the need for a background model. This only has a moderate effect on the expected sensitivities (which currently assume that every signal hypothesis predicting more than 3 events can be excluded at 95\% confidence level). In figure~\ref{fig:dvmet_excl_nobs5} we show the projected sensitivities for the case that both modified analyses lead to 5 observed events and that no background subtraction is attempted, while the existing analysis is assumed to remain background-free even at $300 \, \mathrm{fb^{-1}}$. One can see that even with these very conservative assumptions the proposed modifications make it possible to probe parameter regions at small LLP masses inaccessible to the existing analysis.

Of course searches with unknown backgrounds are by definition unable to discover evidence for exotic long-lived particles. Nevertheless, by setting strong exclusion bounds on the parameter space of such models, they can help to establish the most relevant parameter regions, which can then be targeted with different search strategies.

\section{Conclusion}
\label{sec:conclusions}

Long-lived particles with masses at the GeV scale are a generic prediction of many dark sector models.  However, existing LHC searches for displaced vertices are typically motivated by TeV-scale physics and optimised for LLPs with masses above 100~GeV, which substantially reduces their sensitivity to lighter long-lived particles. In this paper we have pointed out this gap in the current search programme for long-lived particles at the LHC and proposed possible strategies to close it.

As a specific example, we have studied the sensitivity of an ATLAS search for displaced vertices and missing energy to two qualitatively different dark sector models that contain light LLPs and are at present largely unconstrained. The first example is a strongly-interacting dark sector giving rise to dark showers that contain both long-lived and stable dark mesons. The second model is a Higgsed dark sector in which the decay of the dark Higgs boson is loop-suppressed. 
In both cases the production of DM particles at the LHC is generically accompanied by long-lived particles, giving rise to displaced vertices in association with missing energy. The dark shower signal and the dark Higgs signal differ in their typical LLP multiplicity and kinematics. An additional qualitative difference arises from the fact that the dark Higgs decays predominantly into bottom quarks, which hadronise to long-lived $B$ mesons.

Despite these differences, we have found that in both models the cuts of the existing ATLAS analysis result in tiny signal efficiency if the LLP has a mass at the GeV-scale. In particular, the requirement that the impact parameter of all tracks included in the vertex reconstruction be larger than 2~mm biases the reconstructed vertex mass to values well below the original LLP mass. Thus, a large number of vertices fail the mass cut, in particular for LLPs with small mass and/or short decay length. Another limiting factor on the sensitivity turns out to be the cut on the number of tracks, particularly for LLPs decaying into heavy quarks.

As a result, we have found that the existing search does not make an exclusion in the dark shower model and only a marginal exclusion in the dark Higgs model. Hence, we have proposed two modifications to the vertex reconstruction and analysis cuts: the {\it `relaxed cuts'} analysis relaxes the cut on the number of tracks and the vertex mass while remaining background-free; the {\it `relaxed $d_0$'} analysis requires only two of the tracks joined to a vertex to have impact parameter $d_0>2$~mm. The {\it `relaxed cuts'} analysis is particularly useful for  LLP masses below 20~GeV, while the {\it `relaxed $d_0$'} analysis enhances the signal efficiency at larger masses and short decay length by more than an order of magnitude. We have demonstrated the impact of these improvements by deriving projected exclusion limits for these modifications in the dark shower model and the dark Higgs model (see figure~\ref{fig:dvmet_excl}).
For both models we find that searches for displaced vertices and missing energy are sensitive to parameter regions inaccessible with prompt signatures and complementary to other types of DM searches.

We note that both of the dark sector models that we consider differ from most SUSY-inspired LLP models in that the LLPs are not necessarily produced in pairs, making searches for a single displaced vertices particularly important. Nevertheless, there is a non-negligible probability for several LLPs to be produced in the same event, in particular in the dark shower model. Hence, searches for pairs of displaced vertices~\cite{Aaboud:2018aqj,Sirunyan:2018vlw,Aaboud:2019opc,Aad:2019xav} may also offer some sensitivity to these models. The combination of these different search strategies furthermore offers the exciting possibility to map out the LLP multiplicity and thereby discriminate different LLP models.

Finally, we emphasize that much stronger exclusion limits could potentially be obtained by combining the two modifications that we propose, or by extending the search window to even smaller vertex masses and number of tracks. The combination of these different approaches will maximise the discovery potential of the LHC.

\acknowledgments
We thank Juliette Alimena, James Beacham, Lawrence Lee, Christian Ohm, Simone Pagan Griso and Jos\'e Zurita for discussions. This  work  is  funded  by  the  Deutsche Forschungsgemeinschaft (DFG) through the Collaborative Research Center TRR 257 ``Particle Physics Phenomenology after the Higgs Discovery'' under Grant  396021762 -- TRR 257 and the Emmy Noether Grant No.\ KA 4662/1-1.

\bibliographystyle{JHEP_improved}
\bibliography{simp}

\end{document}